\documentclass{IEEEtran}
\usepackage{cite}
\usepackage{amsmath,amssymb,amsfonts}
\usepackage{graphicx}
\usepackage{textcomp}
\usepackage[breaklinks]{hyperref}
\usepackage{amsmath,amssymb,amsmath,amsthm,bm,graphicx,xcolor}
\usepackage{empheq}
\usepackage{algorithm}
\usepackage{algpseudocode}
\usepackage{xfrac}
\usepackage{cuted}
\usepackage{mathrsfs}
\usepackage{float}
\usepackage{graphicx}
\usepackage{xcolor}
\usepackage{soul}
\definecolor{PineGreen}{RGB}{1, 121, 111}
\begin{document}
\title{Time-Domain Topology Optimization of Power Dissipation in Dispersive Dielectric and Plasmonic Nanostructures}
\author{Johannes Gedeon, Izzatjon Allayarov, Antonio Cal{\`a} Lesina, and Emadeldeen Hassan
\thanks{
This work was supported by the Deutsche Forschungsgemeinschaft (DFG, German Research Foundation) under Germany’s Excellence Strategy within the Cluster of Excellence PhoenixD (EXC 2122, Project ID 390833453), and within the Research Grant CA 2763/2-1 (Project ID 527470210). The computing time was granted by the Resource Allocation Board and provided on the supercomputer Lise and Emmy/Grete at NHR@ZIB and NHR@Göttingen as part of the NHR infrastructure. The calculations for this research were conducted with computing resources under the project nip00059. J.G., A.C.L. and E.H. formulated the research question, J.G. developed the theory, implemented the code, performed the study and wrote the manuscript. I.A. provided the Mie analysis section. A.C.L and E.H. supervised the work and revised the manuscript.}
\thanks{Johannes Gedeon, Izzatjon Allayarov and Antonio Cal{\`a} Lesina are with
the Hannover Centre for Optical Technologies, the Institute for Transport and Automation Technology (Faculty of Mechanical Engineering), and the Cluster of Excellence PhoenixD,
Leibniz University Hannover, 30167 Hannover, Germany (e-mail: johannes.gedeon@hot.uni-hannover.de; antonio.calalesina@hot.uni-hannover.de).}
\thanks{Emadeldeen Hassan is with the Department of Applied Physics and Electronics, Umeå University, SE-901 87 Umeå, Sweden.}}

\maketitle

\begin{abstract}
We present a density-based topology optimization scheme for locally optimizing the electric power dissipation in nanostructures made of lossy dispersive materials. By using the complex-conjugate pole-residue (CCPR) model, we can accurately model any linear materials' dispersion without limiting to specific material classes. We incorporate the CCPR model via auxiliary differential equations (ADE) into Maxwell's equations in the time domain, and formulate a gradient-based topology optimization problem to optimize the dissipation over a broad spectrum of frequencies. To estimate the objective function gradient, we use the adjoint field method, and explain the discretization and integration of the adjoint system into the finite-difference time-domain (FDTD) framework. Our method is demonstrated using the example of topology optimized spherical nanoparticles made of Gold and Silicon with an enhanced absorption efficiency in the visible-ultraviolet spectral range. In this context, a detailed analysis of the challenges of topology optimization of plasmonic materials associated with a density-based approach is given.
\end{abstract}


\section{Introduction}
\label{sec:introduction}
\IEEEPARstart{T}{opology} optimization (TopOpt) in nanophotonics has proven to be a powerful method to generate novel nanostructures with desired optical functionalities. 
Its inverse design approach enables the development of optical components that far surpass those created through human intuition and physical first principles in terms of efficiency. 
Topology optimization commonly refers to a density-based approach and the use of the adjoint method to enable a gradient-based optimization\cite{Bendsoe2004,Jensen}. 
The density is mapped to the material's distribution in space, and is iteratively updated to maximize a given objective function. 
By using the adjoint method, the gradients are calculated with only two simulations. 
Over the past decade, it has been applied to various engineering challenges, from developing achromatic metalenses \cite{metalens} and broadband (subwavelength-) antennas~\cite{Nomura07Structural,Erentok11,wang17Antenna, Gedeon2023, Hassan:22}, to single-photon emitters for quantum computing \cite{photonemitter}, nonlinear multiplexer\cite{NL1}, multilayer waveguides transitions \cite{transition}, and small-scale particle accelerators \cite{accelerator}.\par 
Most of these optimization objectives are directly or indirectly linked to the electric energy, and typically operate in a frequency regime in which the optical material dispersion and losses are negligible. However, towards higher frequencies and with the requirement of broadband performance, the assumption of a dispersionless material cannot be made. Dissipation (or absorption), like energy itself, is a fundamental quantity of the Poynting theorem and intrinsically linked to dispersion by thermodynamics laws \cite{landau1995electrodynamics}. 
Absorption is commonly optimized implicitly by tackling reflection and transmission in the far-field \cite{WANG2023129526,Miller}, or based on the physical properties of specific combinations of materials,
 such as metal-dielectric-metal and multi-layered architectures \cite{SolarAbsorber,Jiang2021,Shrestha2018, LIU2021107575}, photonic crystals \cite{Starczewska2024, Rinnerbauer:14, Dong:23}, or incorporation of plasmonic nanoparticles into the absorbing material \cite{Omrani2022, antoniosSolar}. It is timely to conduct an in-depth examination of this physical quantity in the context of topology optimization for dispersive nanostructures.\par
In this work, we present a topology optimization method for locally optimizing the electric power dissipation for arbitrary dispersive materials. Contrary to conventional methods, our approach is explicit because the design region overlaps/coincides with the region where the optimization objective is specified. This opens the door to solving novel design problems where the dissipation (or conversion of energy) in the structure itself is the objective of interest, with possible implications in solar energy conversion in (thermo-) photovoltaics \cite{solarReview}, thermo-nanophotonics \cite{Baffou2020,Zograf2021}, radiotherapy and thermal ablation \cite{Cilla2020}, photodetection \cite{Lin2014, Chen2012,Shrekenhamer}, absorptive filters and sensors \cite{Bruckner:13, filter1, filter2}, or plasmonic devices with a reduced power loss \cite{anapolePaper}.\par 
We use the adjoint method, relying on the formulation of Maxwell’s equations in the time domain. Time-domain topology optimization increasingly attracts attention as it is well suited to tackle a variety of special optimization problems, such as the optimization of dynamic and transient effects, pulse shaping, and time-varying materials \cite{Gedeon2023, Yasuda2021,Baxter:23,Tang2023}. Since a time-domain approach allows the entire spectral content of interest to be incorporated into the time-dependent source of excitation, the method allows us to target an optimized performance over a broad continuous frequency spectrum. Furthermore, dispersive materials can be simulated with the finite-difference time-domain (FDTD) method, e.g., via the auxiliary differential equation (ADE) approach \cite{Greene:06,6740092,8327445,material}, and can therefore be optimized across a broad spectrum. We recently demonstrated such an approach by optimizing the field confinement in dielectric and plasmonic nanostructures based on the complex-conjugate pole-residue (CCPR) model \cite{Gedeon2023}. The CCPR model is not restricted to a certain physical model, as it allows the fitting of any dielectric function \cite{1603585, material}.  We build on these results and integrate this model in our equations to enable the maximization/minimization of the dissipation for arbitrary materials.\par 
Starting from the Poynting theorem, we find the expression of the instantaneous electric power dissipation of lossy dispersive media described by the CCPR model (Sec.~\ref{sec:PowerDissipation}). The time-average of this expression serves as our objective function, and we present the corresponding adjoint scheme in Sec.~\ref{sec:AdjointMethod} to enable a gradient-based optimization. The discretization of the adjoint system using the FDTD method is presented in Sec.~\ref{sec:FDTDImplementation}. 
A brief overview of the common techniques in density-based topology optimization is given in Sec.~\ref{sec:TopOpt_techniques}. In this context, we provide a heuristic procedure to determine an artificial damping term, which has a significant impact on the optimization convergence when plasmonic materials are considered. 
As a study case and to demonstrate our method, we chose the topology optimization of spherical Silicon and Gold nanoparticles with an enhanced absorption efficiency in the visible-ultraviolet regime. The results are presented in Sec.~\ref{sec:Results}, including an assessment of the success and challenges associated with our optimization method.

\section{Instantaneous electric power dissipation density in lossy dispersive media}\label{sec:PowerDissipation}
In this section, we present the expression for the instantaneous electric power dissipation density based on the CCPR model. The time-averaged dissipation is defined later as our objective function. Given that, we derive the time-domain adjoint scheme to maximize/minimize this quantity within the topology optimization framework in Sec.~\ref{sec:AdjointMethod}. In the following, we consider an isotropic and non-magnetic medium whose dielectric function can be described by CCPR poles with $e^{\,j\omega t}$ time-dependency as
\begin{equation}\label{Eq:eps_CCPR}
\varepsilon(\omega)=\varepsilon_{\infty}+\frac{\sigma}{j \omega \varepsilon_0}+\sum_{p=1}^{P}\left(\frac{c_{p}}{j \omega-a_{p}}+\frac{c_{p}^*}{j \omega-a_{p}^*}\right).
\end{equation}
We note that the coefficients $a_p$ and $c_p$ themselves can be complex, and $*$ represents the complex conjugation. The model can accurately describe the complex permittivity of any lossy dispersive material as long as the number of poles $P$ is chosen appropriately. By the proper selection of the coefficients, this model incorporates
all the standard and advanced dispersion models commonly used, such as Debye, Drude (+ critical points), and (modified) Lorentz. \cite{material}. Using the auxiliary differential equation method, the full system of the time-domain Maxwell's equations in a source-free region reads
\begin{subequations}\label{Eq:Maxwells}
  \begin{empheq}[]{align}
-\nabla \times \mathbf{H} + \varepsilon_{0}\varepsilon_{\infty}\partial_{t}\mathbf{E} + \sigma \mathbf{E} + 2\sum_{p=1}^{P}\Re\left\{\partial_{t}\mathbf{Q}_{p}\right\} &= 0,\\
\forall p \in \{1, \ldots, P\}: \partial_{t}\mathbf{Q}_{p}-a_{p}\mathbf{Q}_{p}- \varepsilon_{0}c_{p}\mathbf{E} &= 0, \\[8pt]
\mu_0 \partial_t \mathbf{H}+\nabla \times \mathbf{E} &=0,
\end{empheq}
\end{subequations}
where the complex auxiliary fields $\mathbf{Q}_{p}$ must be computed for all poles $p \in \{1,..., P\}$. To derive the expression for the instantaneous electric power dissipation based on the CCPR model, we follow a similar procedure as the one presented in Ref.~\cite{Shin:12}, which is therein limited to Lorentz media only. We start from the general Poynting theorem in a source-free \mbox{domain $\Omega_{\mathrm{o}}$}\cite{Nunes:11},
\begin{equation}\label{Eq:PoyntingTheorem_1}
-\oint_{\partial\Omega_{\mathrm{o}}}(\mathbf{E} \times \mathbf{H}) \cdot \mathrm{d} \mathbf{a}=\int_{\Omega_{\mathrm{o}}}\left(\mathbf{E} \cdot \frac{\partial \mathbf{D}}{\partial t}+\mathbf{H} \cdot \frac{\partial \mathbf{B}}{\partial t}\right) \mathrm{d}^3 r,
\end{equation}
with the ansatz
\begin{equation}\label{Eq:PoyntingTheorem_splitting}
\mathbf{E} \cdot \frac{\partial \mathbf{D}}{\partial t} = \frac{\partial u_e}{\partial t}+ q_e.
\end{equation}
Here, $u_e(t)$ represents the instantaneous electric energy density, and $q_e(t)$ is the instantaneous electric dissipation density, respectively. Using the relation $\partial_t \mathbf{D}=\nabla \times \mathbf{H}$, we can calculate the product in Eq.~(\ref{Eq:PoyntingTheorem_splitting}) and separate the terms according to the ansatz on the right-hand side (Supplementary Material, Sec.~II). As a result, the electric power dissipation density can be identified as 
\begin{equation}\label{Eq:InstantanousDissipation}
q_e(t) = \sigma \mathbf{E}^{2} + 2\sum_{p=1}^P \Re\left\{\frac{(\partial_{t}\mathbf{Q}_p)^2}{\varepsilon_{0}c_p}\right\}.
\end{equation}
We note that this equation is valid for any time-varying electric field $\mathbf{E}(t)$. To evaluate the dissipation over an arbitrary spectral range of frequencies in the steady state (Appendix~\ref{app:Dissip_freq_domain}), we now assume a time harmonic field $\mathbf{E}(t):= \Re\{\mathbf{\hat{E}}_0 e^{j\omega t}\}$, where the amplitude $\mathbf{\hat{E}}_0$ itself is frequency independent. In that case, the auxiliary fields from Eq.~(\ref{Eq:Maxwells}b) take the form 
\begin{equation}\label{Eq:Q_fields_freq}
\mathbf{Q}_{p}(t)= \,\frac{1}{2}\left(\frac{\varepsilon_0 c_{p}}{j \omega-a_{p}} \hat{\mathbf{E}}_{0} e^{j \omega t}+\frac{\varepsilon_0 c_{p}}{-j \omega-a_{p}} \hat{\mathbf{E}}_{0}^{*} e^{-j \omega t}\right).
\end{equation}
 Substituting this expression into the Eq.~(\ref{Eq:InstantanousDissipation}) and averaging with respect to time, the product terms containing factors $e^{\pm 2j\omega t}$ give zero \cite{Shin:12, landau1995electrodynamics}. This leaves the equation of the time-averaged electric power dissipation,
\begin{equation}\label{Eq:FreqDissipation}
\bar{q}_e(\omega)=\left(\frac{1}{2}\sigma + \sum_{p=1}^{P}\Re\left\{\frac{\omega^{2} \varepsilon_{0} c_{p}}{(j\omega - a_{p})(-j\omega - a_{p})}\right\}\right)|\hat{\mathbf{E}}_0|^{2}.
\end{equation}
This expression can also be referred to as the \textit{dc component} of the instantaneous electric power dissipation. The result is in agreement with the model-independent expression derived by Landau and Lifschitz \cite{landau1995electrodynamics},
\begin{equation}\label{Eq:Landau_dissipation}
\bar{q}_e= \frac{1}{2}\varepsilon_{0}\omega \varepsilon^{\prime\prime}|\mathbf{\hat{E}}_0|^{2},   
\end{equation}
considering the convention $\varepsilon= \varepsilon^{\prime} - j\varepsilon^{\prime\prime}$ of the permittivity.
\section{Methods}\label{sec:Methods}
\subsection{Adjoint Scheme}\label{sec:AdjointMethod}
Here, we present the time-dependent adjoint scheme to optimize the power dissipation in dispersive media within the density-based topology optimization framework. The formulation builds on the method proposed in Ref.~\cite{Gedeon2023}. A comprehensive derivation is provided in the Supplementary Material, Sec. III. \newline 
Density-based topology optimization (TopOpt) for inverse design refers to an iterative design process that allows us to optimize the distribution of a given material in a specified domain in order to optimize a certain objective function \cite{Bendsoe2004}. The method requires the description of the material in the design domain as a spatial density \mbox{distribution $0\leq \rho(\bm{r}) \leq 1$} which is mapped to the material’s properties, such as the permittivity given in Eq.~(\ref{Eq:eps_CCPR}) in our case, and consequently describes the topological shape of the photonic device as the density converges to
a binary design.\par
\begin{figure}[t!]
    \centering
    \includegraphics[width=0.42\linewidth]{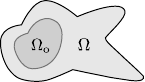}
    \caption{Schematic illustration of the sets $\Omega$ and $\Omega_{\mathrm{o}}$. The objective \mbox{function $F$} is defined on a subset $\Omega_{\mathrm{o}}$ (observation region) lying in the design region $\Omega$. The design material can vary in both regions during the optimization.}
    \label{fig:Sets}
\end{figure}

 
In the following, we consider a design region $\Omega$ and define the objective function on a subset $\Omega_{\mathrm{o}} \subseteq \Omega$ as
\begin{equation}\label{Eq:ObjectiveDissipation}
F := \frac{1}{T}\int_{\Omega_{\mathrm{o}} \times I} q_{e}(t,\rho(\bm{r})) \mathrm{d} t\mathrm{d}^3r.   
\end{equation}
This equation represents the time-averaged electric power loss in a subregion $\Omega_{\mathrm{o}}$ lying in a material-interpolated design over a time interval $I=[0,T]$. $T$ denotes the time period until the fields are decayed after the excitation of a light pulse, which is injected outside the domain $\Omega$. If $\Omega \equiv \Omega_{\mathrm{o}}$ and the design is considered to be entirely surrounded by a non-absorbing medium, the objective is equivalent to the average power transferred from the light pulse to the nanostructure.  
Fig.~\ref{fig:Sets} shows a schematic representation of the sets $\Omega$ and $\Omega_{\mathrm{o}}$, which we refer to when deriving the gradients of the objective with respect to the density on subsets. 
Following a linear interpolation scheme of the parameters $\varepsilon_{\infty}$, $\sigma$ and the pole terms of the CCPR model, Eq.~(\ref{Eq:eps_CCPR}), between background and design material, the density-dependent dissipation term reads
\begin{equation}\label{Eq:InstantanousDissipation_densitydep}
q_e(t,\rho) = \sigma(\rho) \mathbf{E}^{2} + 2\sum_{i=1}^{2}\kappa^{(i)}(\rho)\sum_{p=1}^P \Re\left\{\frac{(\partial_{t}\mathbf{Q}_{p}^{(i)})^2}{\varepsilon_{0}c_{p}^{(i)}}\right\},
\end{equation}
where $\kappa^{(1)}(\rho):=1-\rho$ and $\kappa^{(2)}(\rho):= \rho$. The indices $i=1,2$ represent the background and design material, respectively. We expand the linear mixed parameter $\sigma(\rho)$ by an additional term, such that
\begin{equation}\label{Eq:sigma_mix0}
\sigma(\rho): = \sum_{i \in\{1,2\}}\kappa^{(i)}(\rho)\sigma^{(i)} + \rho (1- \rho) \gamma.
\end{equation}
Here,  $\gamma$ is an artificial damping parameter. As we will see, this parameter plays a crucial role for the convergence based on the objective we have chosen, as it is directly incorporated into Eq.~(\ref{Eq:InstantanousDissipation_densitydep}) and its gradients with respect to the density. 
We describe our choices and its impact on the example of Gold in detail in Sec.~\ref{sec:dampingterm}.\newline
Defining two material-dependent fields $\mathbf{f}^{(i)}$, $i=1,2$, as 
\begin{equation}
\mathbf{f}^{(i)}:=\varepsilon_{0}\varepsilon_{\infty}^{(i)}\partial_{t}\mathbf{E} + \sigma^{(i)} \mathbf{E} + 2\sum_{p=1}^{P^{(i)}}\Re\left\{\partial_{t}\mathbf{Q}_{p}^{(i)}\right\},
\end{equation}
the Maxwell's Eqs.~(\ref{Eq:Maxwells}) are changed according to their density dependency in $\Omega$ to
\begin{subequations}\label{Eq:Forward}
\begin{align}
-\nabla \times \mathbf{H} +\sum_{i=1}^{2}\kappa^{(i)}(\rho)\,\mathbf{f}^{(i)} + \rho (1- \rho) \gamma \,\mathbf{E} &= 0,\\
\substack{
  \text{For } i =1,2 \text{ and }\\
  \forall p\, \in \{1, \ldots, P^{(i)}\}
} \quad \partial_{t}\mathbf{Q}_{p}^{(i)}-a_{p}^{(i)}\mathbf{Q}_{p}^{(i)}- \varepsilon_{0}c_{p}^{(i)}\mathbf{E} &= 0, \\[8pt]
\mu_0 \partial_t \mathbf{H}+\nabla \times \mathbf{E} &=0.
\end{align}
\end{subequations}
The auxiliary fields $\mathbf{Q}_{p}^{(i)}$ must be computed for both materials $i=1,2$ and each corresponding pole $p \in \{1,..., P^{(i)}\}$. We refer to this system of equations in the following as \textit{forward simulation}, in which the objective function is measured. \par 
To update the design using a gradient-based optimization method, we need the gradient information of the objective function with respect to the density. Therefore, we perform an additional \textit{adjoint simulation}, which differs from the forward one only in terms of excitation of the physical system. Based on the interpolation scheme and the definition of our objective, we introduce adjoint sources that are injected into the observation region $\Omega_{\mathrm{o}}$. They consist of an electric source term,
\begin{equation}\label{Eq:AdjointSourceTerms1}
\mathbf{S}_{E}:= \phantom{-}2\,T^{-1}\sigma(\rho)\overleftarrow{\mathbf{E}},
\end{equation}
and auxiliary adjoint source terms,  
\begin{equation}
\mathbf{S}_{\partial_{\tau}Q_{p}}^{(i)}:=\phantom{-}2 \,T^{-1} \partial_{\tau}\overleftarrow{\mathbf{Q}}_{p}^{(i)},
\end{equation}
for $i=1,2$ and $\forall p \in \{1, \ldots, P^{(i)}\}$.
Here, \mbox{$\tau=T-t$} denotes the transformed variable after time reversal, and the symbol "$\overleftarrow{}$" marks the time reversal of the corresponding fields calculated during the forward simulation. We again define two material-dependent fields $\tilde{\mathbf{f}}^{(i)}$, $i=1,2$ as
\begin{equation}
\tilde{\mathbf{f}}^{(i)}:=\varepsilon_{0}\varepsilon_{\infty}^{(i)}\partial_{\tau}\tilde{\mathbf{E}} + \sigma^{(i)} \tilde{\mathbf{E}} + 2\sum_{p=1}^{P^{(i)}}\Re\left\{\partial_{\tau}\tilde{\mathbf{Q}}_{p}^{(i)}\right\}.
\end{equation}
Then, the adjoint system on $\Omega_{\mathrm{o}}$ reads
\begin{subequations}\label{Eq:Adjoint}
\begin{align}
-\nabla \times \tilde{\mathbf{H}} +\sum_{i=1}^{2}\kappa^{(i)}(\rho)\,\tilde{\mathbf{f}}^{(i)} + \rho (1- \rho) \gamma \,\tilde{\mathbf{E}} &= \mathbf{S}_{E},\\
\substack{
  \text{For } i =1,2 \text{ and }\\
  \forall p\, \in \{1, \ldots, P^{(i)}\}
} \;\;\partial_{\tau}\tilde{\mathbf{Q}}_{p}^{(i)}-a_{p}^{(i)}\tilde{\mathbf{Q}}_{p}^{(i)}- \varepsilon_{0}c_{p}^{(i)}\tilde{\mathbf{E}} &= \mathbf{S}_{\partial_{\tau}Q_{p}}^{(i)}, \\[8pt]
\mu_0 \partial_{\tau} \tilde{\mathbf{H}}+\nabla \times \tilde{\mathbf{E}} &=0.
\end{align}
\end{subequations}
Since the objective $F$ in Eq.~(\ref{Eq:ObjectiveDissipation}) is defined on the subset $\Omega_{\mathrm{o}}$ only, no adjoint sources will be injected into the domain $\Omega \setminus \Omega_{\mathrm{o}}$. In that case, the right-hand side of the system of equations~(\ref{Eq:Adjoint}) is 0. The adjoint fields $\{\tilde{\mathbf{E}}$, $\tilde{\mathbf{Q}}_{p}^{(i)}\}$ together with the forward fields $\{\mathbf{E}$, $\mathbf{Q}_{p}^{(i)}\}$ are used to calculate the gradient of the objective function $F$ with respect to the density, 
\begin{equation}\label{Eq:GradientSuperposition}
\nabla_{\rho}F:= \mathcal{A}_{|\rho\in\Omega}+ \mathcal{B}_{|\rho\in\Omega_{\mathrm{o}}},
\end{equation}
which is a sum of two gradient terms defined on each set $\Omega$ and $\Omega_{\mathrm{o}}$ illustrated in Fig.~\ref{fig:Sets}, namely
\begin{subequations}\label{Eq:GradientsTimeReversal_Diss}
\begin{align} 
\mathcal{A}_{|\rho\in\Omega}:=&\;-\int_{I}\varepsilon_{0}(\mathrm{d}_{\rho}\varepsilon_{\infty})\partial_{\tau}\tilde{\mathbf{E}} \cdot \overleftarrow{\mathbf{E}}  \\
&-\int_{I}(\mathrm{d}_{\rho}\sigma) \tilde{\mathbf{E}} \cdot \overleftarrow{\mathbf{E}} \notag \\
&-\int_{I}\sum_{i=1}^{2}\sum_{p=1}^{P^{(i)}}2(\mathrm{d}_{\rho}\kappa^{(i)})\partial_{\tau}\tilde{\mathbf{E}}\cdot\Re\left\{\overleftarrow{\mathbf{Q}}_{p}^{(i)}\right\}, \notag\\[8pt]
 \mathcal{B}_{|\rho\in\Omega_{\mathrm{o}}}:=&\;\;\;\;\,\frac{1}{T}\int_{I} (\mathrm{d}_{\rho}\sigma) \overleftarrow{\mathbf{E}}^{2}\\ \notag
 &+\frac{1}{T}\int_{I} \sum_{i=1}^{2}\sum_{p=1}^{P^{(i)}} 2(\mathrm{d}_{\rho}\kappa^{(i)}) \Re\left\{\frac{(\partial_{\tau}\overleftarrow{\mathbf{Q}}_{p}^{(i)})^2}{\varepsilon_{0}c^{(i)}_p}\right\}. \notag
\end{align}
\end{subequations}

From these equations, we observe simplifications for two important special cases: 
\begin{itemize}
    \item If both background and design material in $\Omega$ do not contain any CCPR poles, i.e., they are both  non-dispersive and their losses are only described by the $\sigma$ term in Eq.~(\ref{Eq:eps_CCPR}), all auxiliary fields vanish. And since $\mathbf{S}_{\partial_{\tau}Q_{p}}^{(i)}\equiv0$, no auxiliary adjoint source will be injected during the adjoint simulation. In addition, the gradients in Eqs.~(\ref{Eq:GradientsTimeReversal_Diss}) do not contain any terms related to the $\mathbf{Q}_{p}^{(i)}$ fields.
    \item If the objective function in Eq.~(\ref{Eq:ObjectiveDissipation}) defined on $\Omega_{\mathrm{o}}$ does not depend on the density itself, i.e., the density does not vary in $\Omega_{\mathrm{o}}$ during the optimization, the derivatives of the material parameters with respect to the density vanish in $\Omega_{\mathrm{o}}$. And since $\forall\rho\in\Omega_{\mathrm{o}}$: $\mathrm{d}_{\rho}\varepsilon_{\infty}=\mathrm{d}_{\rho}\sigma=\mathrm{d}_{\rho}\kappa^{(i)}=0$, it follows that $\mathcal{B}_{|\rho\in\Omega_{\mathrm{o}}}\equiv0$, i.e., the total gradient reduces to $\nabla_{\rho}F:= \mathcal{A}_{|\rho\in\Omega\setminus \Omega_o}$, in agreement with the results presented in Refs.~\cite{Gedeon2023, transition}.
\end{itemize}

\subsection{Density-based topology optimization techniques}\label{sec:TopOpt_techniques}
\subsubsection{Filtering and Projection}
In density-based topology optimization, it is common to apply a filter and projection on the density $\rho$. It can serve the purpose of eradicating the appearance of single-pixel features, introducing a weak sense of length-scale into the design, or curing the self-penalization issue when optimizing lossy structures~\cite{Bourdin01Filters, patch2014, Wang2011OnPM, Hassan14Topology}. The transformed form $\bar{\tilde{\rho}}$ describes the material interpolation and is also incorporated into the objective function Eq.~(\ref{Eq:ObjectiveDissipation}) in our case.\newline
At each iteration, we average over a neighborhood around the density points $\rho(\bm{r})$ by applying the filter operator
\begin{equation}\label{FilterFunction}
\tilde{\rho}(\bm{r})=(\mathscr{F}\rho)(\bm{r}) := \frac{\int_{\mathcal{B}_{R}(\bm{r})} w\left(\bm{r}, \bm{r}^{\prime}\right)\rho(\bm{r}^{\prime})\,\mathrm{d}^3r^{\prime}}{\int_{\mathcal{B}_{R}(\bm{r})} w\left(\bm{r}, \bm{r}^{\prime}\right)\,\mathrm{d}^3r^{\prime}}.
\end{equation}
$\mathcal{B}_{R}(\bm{r})$ describes a sphere with radius $R$ around $\bm{r}$, and $w\left(\bm{r}, \bm{r}^{\prime}\right)$ is a weighting function defined as
\begin{equation}
w\left(\bm{r}, \bm{r}^{\prime}\right)=R-\left|\bm{r}-\bm{r}^{\prime}\right|.
\end{equation}
Subsequently, we project the filtered density by the smoothed Heaviside function
\begin{equation}\label{Projection}
\bar{\tilde{\rho}}(\bm{r})=\mathscr{P}(\tilde{\rho}(\bm{r})):=\frac{\tanh (\beta \eta)+\tanh \left(\beta\left(\tilde{\rho}(\bm{r})-\eta\right)\right)}{\tanh (\beta \eta)+\tanh (\beta(1-\eta))}.
\end{equation}
The parameter $\eta$ determines  the threshold, and the parameter $\beta$ controls the sharpness of the projection. For $\beta \rightarrow \infty $, the projected density is binary, i.e., it has only the values 0 or 1.
The derivative of a functional $F[\rho]$  with respect to the original density $\rho$ can be calculated by using the chain rule together with the Fréchet derivatives of the filtering and projection operators,
\begin{equation}\label{GradientsFiltered}
\frac{\delta F[\rho]}{\delta \rho(\bm{r})}=\frac{\delta F[\bar{\tilde{\rho}}]}{\delta \bar{\tilde{\rho}}(\bm{r})}\mathscr{P}^{\prime} (\tilde{\rho}(\bm{r})) (\mathscr{F}^{\prime}\rho)(\bm{r}).
\end{equation}
The parameter $\beta$ value will be increased during the optimization until the objective does not show any significant change. The measure of non-discreteness $M_{\text{nd}}$ serves as an indicator to verify the convergence of an optimized design to a binary solution ~\cite{Si07},
\begin{equation}\label{Eq:MeasureOfNon-discretness}
M_{\text{nd}}:=100 \% \times V_{\Omega}^{-1}\int_{\Omega} 4 \,\bar{\tilde{\rho}}(\bm{r})\left(1-\bar{\tilde{\rho}}(\bm{r})\right)\mathrm{d}^3r.
\end{equation}
This measure has a maximum if all density points have an intermediate value of 0.5, and it is minimized if the density only consists of the values 0 or 1.
At the end of the optimization, the projected density is mapped to a binary design by a threshold with respect to the parameter $\eta$.\newline
\subsubsection{The artificial damping term $\gamma$ for Gold}\label{sec:dampingterm}
The static conductivity $\sigma$ appearing in the permittivity model Eq.~(\ref{Eq:eps_CCPR}), has been interpolated as 
\begin{equation}\label{Eq:sigma_mix}
\sigma(\rho) = \sum_{i \in\{1,2\}}\kappa^{(i)}(\rho)\sigma^{(i)} + \rho (1- \rho) \gamma. 
\end{equation}
The first term represents the linear interpolation between background and design material. The second term related to $\gamma$ was originally introduced to penalize intermediate densities of dielectric materials \cite{jensen_topology_2005}. However, this artificial damping term has proven to be essential for the optimization of plasmonic materials as well \cite{Hassan:22, Gedeon2023, Andkjaer:10}. The reference \cite{Hassan:22} in particular, highlights the significance of a proper choice of the artificial damping parameter $\gamma$, which has therein been heuristically observed as part of optimizing Silver nanoantennas for field confinement in a non-dispersive observation region.\par
We emphasize that the choice of the value of $\gamma$ is also essential when maximizing the dissipation of a spherical Gold nanostructure in the visible spectral range, as will be investigated in Sec.~\ref{sec:Results}. For wavelengths above $\approx 520\, \mathrm{nm}$, Gold nanoparticles typically show localized surface plasmon resonances. Below this wavelength, however, the interband transition dominates, where the Gold's electrons absorb photons and transition from the filled $d$-band to states above the Fermi level in the $sp$-band, which will effectively results in a higher absorption of light. In the context of density-based topology optimization, we must carefully choose the artificial damping term $\gamma$ to ensure a reasonable behavior of the creation and propagation of plasmons for intermediate density values, and to guarantee the convergence to a binary design as a result. We will therefore only focus on the plasmonic region in the following discussion. Although we present a heuristic method specifically for Gold here, the underlying approach is applicable to plasmonic materials in general. The unit for the artificial damping term $\gamma$ is the same as that for the static conductivity $\sigma$, which we will omit for the sake of better readability here. \par
We can estimate the influence of the damping term on maximizing the dissipation of a spherical Gold structure by analyzing the propagation of surface plasmon polaritons (SPP) along the interface between air and Gold in 1D. In that case, the prerequisites for the formation of surface plasmons are well studied \cite{Maier2007}. The assumption of an infinite flat layer, rather than a complexly shaped surface, represents a drastic simplification. However, it is reasonable as a first approximation if we interpret  resonant localized surface plasmons  of a sphere as the propagation of two counterpropagating surface plasmon waves \cite{Fang_2018}, which later also serves as our initial design. We consider a surface plasmon propagating along $x^{+}$  at the interface of air ($z >0$) and Gold ($z< 0$). Inside Gold, its magnetic and electric field components are described as
\begin{figure}[!t]
\centerline{\includegraphics[width=\columnwidth]{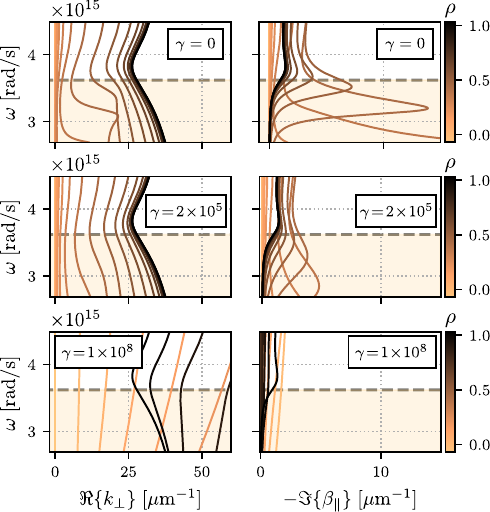}}
\caption{Real part $\Re\{k_{\perp}\}$ (1st column) and negative imaginary part $-\Im\{\beta_{\parallel}\}$ (2nd column) from Eqs.~(\ref{Eq:k_vecs}) for a material-interpolated permittivity $\varepsilon(\bar{\rho})$ of Gold and air. The density $\rho$ was projected according to Eq.~(\ref{Projection}) with $\eta=0.55$ and $\beta=7$.
Each row corresponds to a different artificial damping term $\gamma$ used in the mixing of the static conductivity in Eq.~(\ref{Eq:sigma_mix}). The colored area marks the frequency range in which Gold mainly shows plasmonic behavior. The damping term $\gamma=2\times10^{5}$ has been chosen for the optimization of the Gold nanostructure in Sec.~\ref{sec:Results}.}
\label{fig:kvectors}
\vspace{1cm}
\centerline{\includegraphics[width=.8\linewidth]{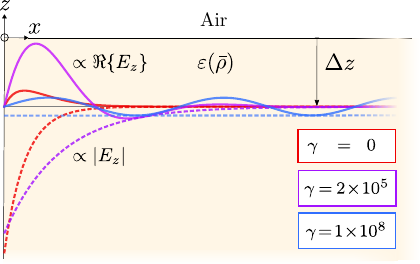}}
\caption{Propagating surface plasmons along the interface of air and the interpolated material $\varepsilon(\bar{\rho})$ for different artificial damping parameters $\gamma$. All plasmons have the same frequency of $\omega=2.8\times 10^{15}$ rad/s, and the density $\rho=0.4$ has been used for the interpolation of the permittivities of air and Gold (with a preceding projection), see Fig.~\ref{fig:kvectors}. The dashed and solid lines represent the magnitude and real part of $E_{z}$ from Eq.~(\ref{Eq:SSP_fields}c) at a depth of $\Delta z=22$ nm within the material, respectively.}
\label{fig:propagation}
\end{figure}
\begin{subequations}\label{Eq:SSP_fields}
\begin{align}
& H_y(x, z)=A \mathrm{e}^{-j \beta_{\parallel} x} \mathrm{e}^{k_{\perp} z}, \\
& E_x(x, z)=j A \frac{1}{\omega \varepsilon_0 \varepsilon} k_{\perp} e^{-j \beta_{\parallel} x} \mathrm{e}^{k_{\perp} z}, \\
& E_z(x, z)=-A \frac{\beta_{\parallel}}{\omega \varepsilon_0 \varepsilon} e^{-j \beta_{\parallel} x} \mathrm{e}^{k_{\perp} z}.
\end{align}
\end{subequations}
 $A$ is an arbitrary amplitude, $\varepsilon$ is the relative permittivity of Gold, and $k_{\perp}$ and $\beta_{\parallel}$ represent the attenuation and the propagation constants of the surface plasmon, respectively, 
 \begin{subequations}\label{Eq:k_vecs}
 \begin{align}
k_{\perp}=&\;\sqrt{\beta_{\parallel}^2-k_0^2 \varepsilon}, \\
\beta_{\parallel}=&\;k_0 \sqrt{\frac{\varepsilon}{1+\varepsilon}} .
\end{align}
\end{subequations}
 Now, we replace the permittivity $\varepsilon$ of Gold in \mbox{Eqs.~(\ref{Eq:SSP_fields}) \& (\ref{Eq:k_vecs})} by the material-interpolated permittivity between air and Gold $\varepsilon(\bar{\rho})$. Here, the density $\bar{\rho}=\mathscr{P}(\rho)$ is projected according to Eq.~(\ref{Projection}) with the parameters $\eta=0.55$ and $\beta=7$. The plot of the values $\Re\{k_{\perp}\}$ and $-\Im\{\beta_{\parallel}\}$ for different density values $\rho$ and damping parameters $\gamma$ is presented in Fig.~\ref{fig:kvectors}. $\Re\{k_{\perp}\}$ describes the wave damping inside Gold, while $-\Im\{\beta_{\parallel}\}$ determines the attenuation along the propagation direction. Fig.~\ref{fig:propagation} schematically illustrates the propagation of SPPs having the same frequency for different artificial damping parameters, just below the interface between air and the interpolated material. For this discussion, we included the projected density $\bar{\rho}$ in the material interpolation, since it is also incorporated into Maxwell's equations in due course during the optimization. A finer tuning of the equidistance between the $\Re\{k_{\perp}\}$ functions for intermediate densities can be achieved by a suitable selection of the projection parameters and the artificial damping parameter. The benefits of this will become clear when interpreting the diagrams in the following.\par 
For $\gamma=0$ (Fig.~\ref{fig:kvectors}, 1st row), the $-\Im\{\beta_{\parallel}\}$ functions are characterized by large values and sharp peaks for intermediate densities in the plasmonic region (marked in color). At these frequencies, the SPPs are strongly confined locally around their excitation point. By considering the corresponding $\Re\{k_{\perp}\}$ values in addition, we can conclude that they are characterized by a deep penetration into the material, which will effectively contribute to the dissipation (Fig.~\ref{fig:propagation}, red). 
The design might retain gray areas throughout the entire optimization process, which contributes to high dissipation around the excitation point, but prevents the plasmons from exploring the topology in its full spatial extent. This can result in an unpredictable performance after the final threshold, where the localized effects for intermediate densities that contributed to the design evolvement are eventually erased.\par Choosing a very large damping term $\gamma=1\times 10^{8}$ instead (Fig.~\ref{fig:kvectors}, 3rd row), eradicates the large and sharp peaks of $-\Im\{\beta_{\parallel}\}$, which allows an extended spatial propagation along the surface of the SPPs, as their decay is weak. However, the penetration into the material is drastically reduced, which can be observed from the large values of $\Re\{k_{\perp}\}$ for the majority of intermediate densities. In this case, the interpolated material tends to have the characteristics of a perfect metal, which makes optimization less sensitive, as a dissipation inside the interpolated material is prevented (Fig.~\ref{fig:propagation}, blue).\par
The value $\gamma=2\times 10^{5}$ (Fig.~\ref{fig:kvectors}, 2nd row) is a compromise. The large peaks of the $-\Im\{\beta_{\parallel}\}$-functions are flattened, such that the SPPs can cover a greater distance along the propagation direction for intermediate densities (Fig.~\ref{fig:propagation}, purple). 
Furthermore, a more uniform equidistant pattern of the intermediate $\Re\{k_{\perp}\}$-functions is given, which is bounded by the curves of pure air ($\rho=0$) and Gold ($\rho=1$). We expect this to result in more stable convergence behaviour during optimization. It is not guaranteed that the design will not retain gray areas, but the dominant influence of highly confined plasmons related to an intermediate density will be significantly reduced compared to the choice $\gamma=0$. Therefore, we can expect an improvement in the convergence towards a binary structure.
 Using an unprojected density $\rho$ for the material interpolation yield a dense clustering close to the curve with $\rho=1$ for the majority of intermediate functions here. By its projection with the parameters $\eta=0.55$ and $\beta=7$, as shown in the figure, a more uniform interpolation was achieved.
We chose these values as our initial projection parameters and the material parameter $\gamma=2\times 10^{5}$ when optimizing the spherical nanoparticle made of Gold in Sec.~\ref{sec:Results}.

\subsection{FDTD Implementation}\label{sec:FDTDImplementation}
\begin{figure*}[!t]
    \centering
    \includegraphics[width=0.8\linewidth]{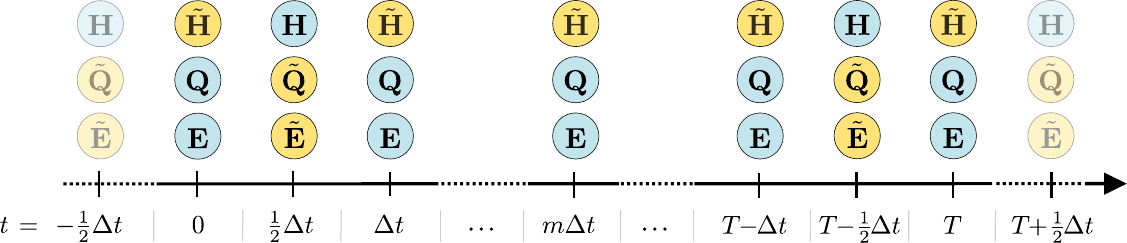}
    \caption{Schematic illustration of the discretized forward and adjoint fields sampled on the timeline within the FDTD framework. The adjoint fields (yellow) are shifted in time by $\Delta t / 2$ compared to the forward fields (blue).}
    \label{fig:Discretization}
\end{figure*}
In the FDTD framework, Maxwell's equations for both forward and adjoint simulation must be discretized on a staggered ``Yee grid" in space and time \cite{Taflove}. The discretization scheme we used is presented in Appendix~\ref{app:discretization}, including the update equations of electric and auxiliary fields. The extension to anisotropic media can be found in the Supplementary Material, \mbox{Sec. III 1-2}. Hereafter, we will limit our discussion to merely explaining time discretization, and present the resulting discretized expression of the gradients in Eqs.~(\ref{Eq:Adjoint}).\newline 
We consider a time interval $[0,T]$, divided into $M+1$ time steps $t_m:=\frac{m}{M}T=m\Delta t$, $m\in \{0,...,M\}$.
The fields for the forward simulation have been discretized in the same way as presented in Ref.~\cite{material}. Therein, \mbox{Eqs.~(\ref{Eq:Maxwells}a) and (\ref{Eq:Maxwells}b)} are discretized at time steps $(m + 1/2)\Delta t$, and Eq.~(\ref{Eq:Maxwells}c) at time steps $m\Delta t$. Consequently, the electric and auxiliary fields are discretized at time steps $m\Delta t$, and the magnetic field at $(m + 1/2)\Delta t$. To avoid a different update scheme for the adjoint system than for the forward system, we discretize both Eqs.~(\ref{Eq:Adjoint}a) and (\ref{Eq:Adjoint}b) at time steps $m\Delta t$, and Eq.~(\ref{Eq:Adjoint}c) at time steps $(m + 1/2)\Delta t$. That means, the sampled adjoint equations are shifted in time by $\Delta t / 2$ compared to the forward equations. As a consequence, both adjoint electric and auxiliary fields must be sampled at time steps $(m + 1/2)\Delta t$, and the adjoint magnetic fields must be sampled at time steps $m\Delta t$. A schematic illustration of the disposition of the discretized fields on the timeline is demonstrated in Fig.~\ref{fig:Discretization}.\newline 
This offset in time, compared to the forward fields, offers two further advantages in terms of numerical accuracy and stability: (1) The injection of the source term $\mathbf{S}_E$ in Eq.~(\ref{Eq:Adjoint}a) does not require further numerical approximation, since it is already sampled at time steps $m\Delta t$ based on the discretization of the forward field $\mathbf{E}$. (2) The gradients terms in Eq.~(\ref{Eq:GradientsTimeReversal_Diss}a) that include the products $\partial_{\tau}\tilde{\mathbf{E}} \cdot \overleftarrow{\mathbf{E}}$ and $\partial_{\tau}\tilde{\mathbf{E}}\cdot\Re\{\overleftarrow{\mathbf{Q}}_{p}^{(i)}\}$ can be computed in the most accurate way when evaluating them at time steps $m\Delta t$, while staying consistent with using the first order forward difference approximation for the time derivative. In return, the adjoint source term $\mathbf{S}_{\partial_{\tau}Q_{p}}^{(i)}$ must be interpolated at time steps $m \Delta t$. Therefore, a centered difference approximation was employed. Compared to the forward equations, the update equations for both the adjoint electric fields and the auxiliary fields include additional terms, coming from the injection of the adjoint sources $\mathbf{S}_E$ and $\mathbf{S}_{\partial_{\tau}Q_{p}}^{(i)}$. 

 Based on this time discretization of forward and adjoint equations, the gradients in Eqs.~(\ref{Eq:GradientsTimeReversal_Diss}) can be computed as follows:
\begin{subequations}\label{Eq:GradientsTimeReversal_Diss_FDTD}
\begin{align}
&\mathcal{A}_{\rho\in\Omega}:=\\[4pt]
-&\sum^{M}_{m=0}\varepsilon_{0}\mathrm{d}_{\rho}\varepsilon_{\infty}\mathbf{E}^{M-m}\cdot\left(\tilde{\mathbf{E}}^{m + \frac{1}{2}} - \tilde{\mathbf{E}}^{m - \frac{1}{2}}\right) \notag \\
 -&\sum^{M}_{m=0}\mathrm{d}_{\rho}\sigma\mathbf{E}^{M-m}\cdot\frac{1}{2}\left(\tilde{\mathbf{E}}^{m + \frac{1}{2}} + \tilde{\mathbf{E}}^{m - \frac{1}{2}}\right)\,\Delta t \notag\\
 -&\sum^{M}_{m=0}\sum_{i=1}^{2}\sum_{p=1}^{P^{(i)}}2\mathrm{d}_{\rho}\kappa^{(i)}\Re\left\{\mathbf{Q}^{M-m, (i)}_{p}\right\}\cdot\left(\tilde{\mathbf{E}}^{m + \frac{1}{2}} - \tilde{\mathbf{E}}^{m - \frac{1}{2}}\right), \notag \\[12pt]
 &\mathcal{B}_{\rho\in\Omega_{\mathrm{o}}}:=\\[4pt]
 &\sum^{M}_{m=0}\frac{\mathrm{d}_{\rho}\sigma}{T}\left(\mathbf{E}^{M-m}\right)^2\Delta t \notag \\
 +&\sum^{M}_{m=0}\sum_{i=1}^{2}\sum_{p=1}^{P^{(i)}}\mathrm{d}_{\rho}\kappa^{(i)}\Re\left\{\frac{\left(\mathbf{Q}^{M-1-m, (i)}_{p} - \mathbf{Q}^{M-m+1, (i)}_{p}\right)^2}{T\,2\,\Delta t\,\varepsilon_{0}\,c^{(i)}_{p}}\right\}. \notag
\end{align}
\end{subequations}

From these equations and the definition of the adjoint source terms, we observe that the electric field $\mathbf{E}$, as well as the auxiliary fields $\mathbf{Q}_{p}^{(i)}$, must be stored in space and time during the forward simulation. This implies a significant increase in memory consumption, the more poles are used for an accurate material description of both background and design materials.\newline
For the spatial discretization across $N$ computational cells, we assign density values $\rho_{n,k}$ on each position of three electric field components $k\in\{1,2,3\}$ at the edges of the $n$th Yee cell. The gradients in Eq.~(\ref{Eq:GradientsTimeReversal_Diss_FDTD}) are calculated per component such that a filtering of the original density as well as of the gradients is essential to obtain a physical meaningful structure \cite{Gedeon2023}.\par 
Algorithm~\ref{alg:adjoint_algorithm} outlines the time-domain gradient-based TopOpt routine using the FDTD method. 
\begin{algorithm}[H] 
\caption{Time-domain gradient-based TopOpt algorithm to maximize/minimize the time-averaged power dissipation using the FDTD method}
\begin{algorithmic}[1] 
\Require Initial density $\rho$, filter radius $R$, initial and maximum penalization parameter $\beta_0$ and $\beta_{\mathrm{max}}$, threshold value $\eta$, the maximum number of iterations $\text{itr}_{\text{max}}$ and time steps $M$.
\Ensure \noindent
\begin{itemize}
\item  $M$ and $\text{itr}_{\text{max}}$ can be determined according to whether the objective function $F$ and the design $\rho$ no longer show any significant change, respectively.
\item  Set a condition $\texttt{cond}_{\,\beta}$ for the increment of $\beta$. \end{itemize}\Statex\hskip\algorithmicindent
\For{$\text{itr}=1, \ldots, \text{itr}_{\text{max}}$}
\State Compute the filtered and projected density $\bar{\tilde{\rho}}$ 
\Procedure{\textsl{Forward Simulation}}{}
\For{$m = 0, \ldots, M$}
\State Compute $\mathbf{H}^{m + \frac{1}{2}}$
\State Store $\mathbf{E}^{m}$ and all $\mathbf{Q}_{p}^{m, (i)}$ 
\State Compute $\mathbf{E}^{m+1}$
\State Compute all $\mathbf{Q}_{p}^{m+1, (i)}$
\State Update objective $F$
\EndFor
\EndProcedure
\Procedure{\textsl{Adjoint Simulation}}{}
\For{$m = 0, \ldots, M$}
\State Compute $\tilde{\mathbf{H}}^{m}$
\State Store current $\tilde{\mathbf{E}}^{m-\frac{1}{2}}$
\State Compute $\tilde{\mathbf{E}}^{m+\frac{1}{2}}$
\State Update $\nabla_{\bar{\tilde{\rho}}} F$ 
\State Compute all $\tilde{\mathbf{Q}}_{p}^{m+\frac{1}{2}, (i)}$
\EndFor
\EndProcedure
    \State Compute $\nabla_{\rho} F$ 
    \State Update $\rho$ 
     \If{$\texttt{cond}_{\,\beta}$ \textit{is true}}
        \State Increase $\beta$ 
    \EndIf
\EndFor
\State Threshold $\bar{\tilde{\rho}}$ with respect to $\eta$ ($\beta \rightarrow \infty$) 
\end{algorithmic}
\label{alg:adjoint_algorithm}
\end{algorithm}

\section{Spherical nanoparticles with enhanced absorption efficiency}\label{sec:Results}
\begin{figure}[t!]
    \centering
    \includegraphics[width=1\linewidth]{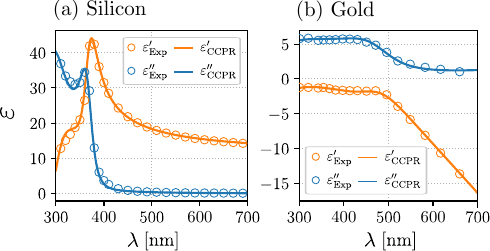}
    \caption{Complex relative permittivity (\mbox{$\varepsilon = \varepsilon^{\prime} - j\,\varepsilon^{\prime\prime}$}) of \mbox{(a) Silicon} and  \mbox{(b) Gold} using the CCPR model in Eq.~(\ref{Eq:eps_CCPR}). Two poles were used for the fit of Silicon, and three poles for Gold. The experimental data is reported in the references~\cite{Si_Schinke, Au_Johnson}, respectively. The corresponding CCPR parameters can be found in Ref.~\cite{Gedeon2023}.}
    \label{fig:Epsilon}
\end{figure}
We applied our method to the inverse design of spherical nanoparticles for an enhanced absorption efficiency in the visible and near ultraviolet regime. To put the generic formulation of the optimization for arbitrary classes of materials under test, we have chosen Gold and Silicon as representatives for metallic and dielectric/semiconductive materials. The particles are embedded in a non-absorbing background medium, which has been chosen as air.\par This optimization problem was selected as our study case due to the following reasons. The absorption efficiency $Q_{\text{abs}}(\omega)$ is proportional to the absorbed power $\mathscr{W}_{\text{abs}}$ of the structure transferred from an incident wave with intensity $I_{\text {inc }}$,
\begin{equation}\label{Eq:Qabs_eq}
Q_{\text{abs}}=\frac{\mathscr{W}_{\text{abs}}}{I_{\text {inc }}\sigma_{\text {geom }}}.
\end{equation}
The geometrical cross section $\sigma_{\text {geom }}=4^{-1}d^2\,\pi$ is determined by the radial expansion of the spherical particle of a diameter $d$, and the absorbed power $\mathscr{W}_{\text{abs}}$ is the integral over the time-averaged electric power dissipation density in Eq.~(\ref{Eq:FreqDissipation}) over the design's volume (considering $\Omega \equiv \Omega_{\mathrm{o}}$). Thus, the value of $Q_{\text{abs}}$ over a broad range of frequencies reflects the success of achieving a broadband performance of dissipation in a direct way. All frequencies of interest can be covered in the source of excitation in our time-domain optimization routine (Sec.~\ref{sec:Parameters}). \newline 
We restrict the design to have a spherical boundary to enable a comparison to analytical solutions using Mie theory and ensure sufficient accuracy of all FDTD simulations. In addition, we use a sphere filled with the design material both as an initial design and as a reference to compare the absorption efficiencies of the final optimized designs. The given setup will highlight the features evolving in 3D, which contribute to the absorption based on different physical effects appearing in dielectric and plasmonic materials. Therefore, we did not enforce explicit manufacturability constraints to allow the design to develop freely in terms of its topological complexity.\newline Moreover, both Silicon and Gold are ideal candidates for the chosen wavelength range of \mbox{300-700 nm} with an imaginary part $\varepsilon^{\prime\prime} \gg 0$ in the dielectric function (Fig.~\ref{fig:Epsilon}), such that an absorbing behavior is expected for both materials in general. The increasing imaginary part towards the violet and near ultraviolet regime for both materials is mainly caused by the direct band transitions in Silicon and the interband transitions in Gold, respectively. The optimization for Gold is considered challenging and represents an ultimate test for broadband optimization in time domain, since both plasmonic effects and penetrating dissipative effects (due to interband transitions) occur in different spectral ranges, respectively. We will see in the following how the structures form topologically to best utilize their complex material properties over the broad spectral range to enable efficient absorption of energy.\par
\begin{figure}[t!]
    \centering
    \includegraphics[width=1\linewidth]{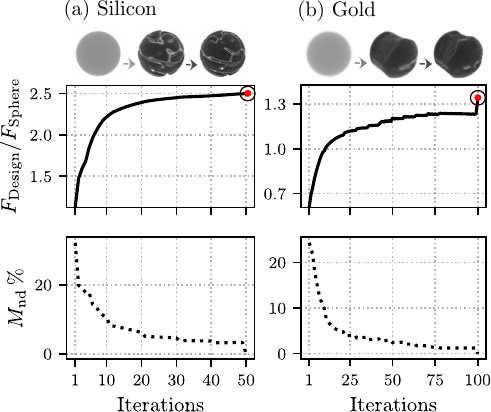}
    \caption{Normalized objective functions (1st row) and measure of non-discreteness (2nd row) per iteration while topology optimizing the (a) Silicon and (b) Gold nanostructures. The normalization constant $F_{\text{Sphere}}$ is the time-averaged power loss calculated for a sphere to whose boundaries the designs were restricted in their expansion. The threshold was applied at the final iteration, marked by a red circle, resulting in a binary design. 
    Before applying the threshold, the measured non-discreteness $M_{\text{nd}}$ of the densities was $3.25\%$  for Silicon and $1.24\%$ for Gold.}
     \label{fig:Convergence}
\end{figure}
\begin{figure*}[!t]
\centerline{\includegraphics[width=1\linewidth]{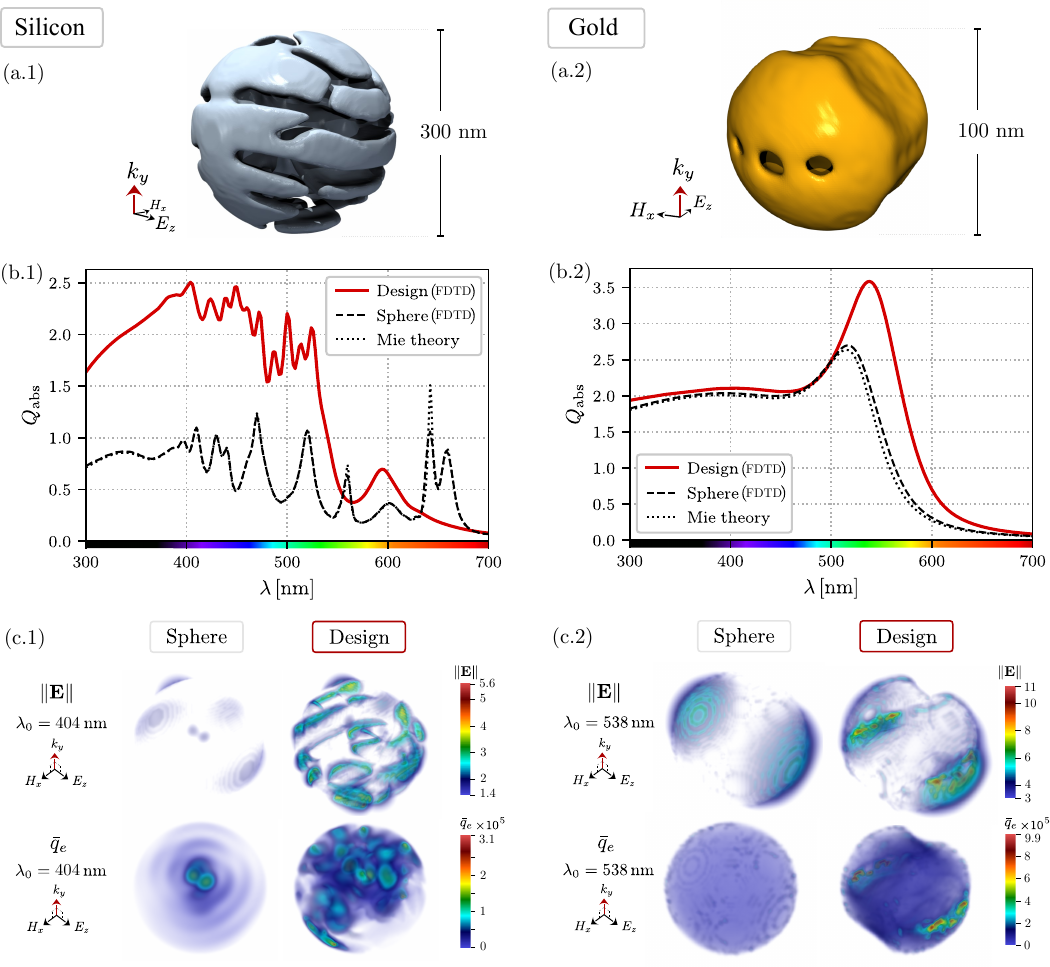}}
\caption{\mbox{(a.1) - (a.2)}: Smoothed illustrations of the optimized nanostructures made of Silicon (left) and Gold (right) for enhanced absorption. Both designs were restricted to evolve within a spherical volume with a fixed diameter during the optimization. The systems are excited by a $z$-polarized plane wave propagating along the $y$-axis. \mbox{(b.1) - (b.2):} Absorption efficiencies of the design (in red) compared to a sphere with the same diameter (black dashed) over the spectral range of interest. They were computed by integrating the time-averaged electric power dissipation density $\bar{q}_e$ in Eq.~(\ref{Eq:FreqDissipation}) over the design's volume, and by using the Fourier-transformed electric fields $\mathbf{E}(\omega)$ from the FDTD simulation. In addition, the absorption efficiencies of a sphere predicted by Mie theory is shown (black dotted). It served as verification of sufficient numerical accuracy before commencing the optimization with identical simulation parameters. \mbox{(c.1) - (c.2):} Spatial distributions of the amplitude of the electric field $\|\mathbf{E}\|$ (1st row) and the time-averaged electric power dissipation density $\bar{q}_{e}$ (2nd row) for the wavelength at which the design has its absorption maximum. The amplitude of the electric field is normalized to that of the incident field. The unit of the dissipation density is given in $\mathrm{W}/\mathrm{m}^{3}$. A comparison with the spatial distributions for a sphere is shown too.}
\label{fig:performance}
\end{figure*}
\begin{figure*}[!t]
\centerline{\includegraphics[width=1\linewidth]{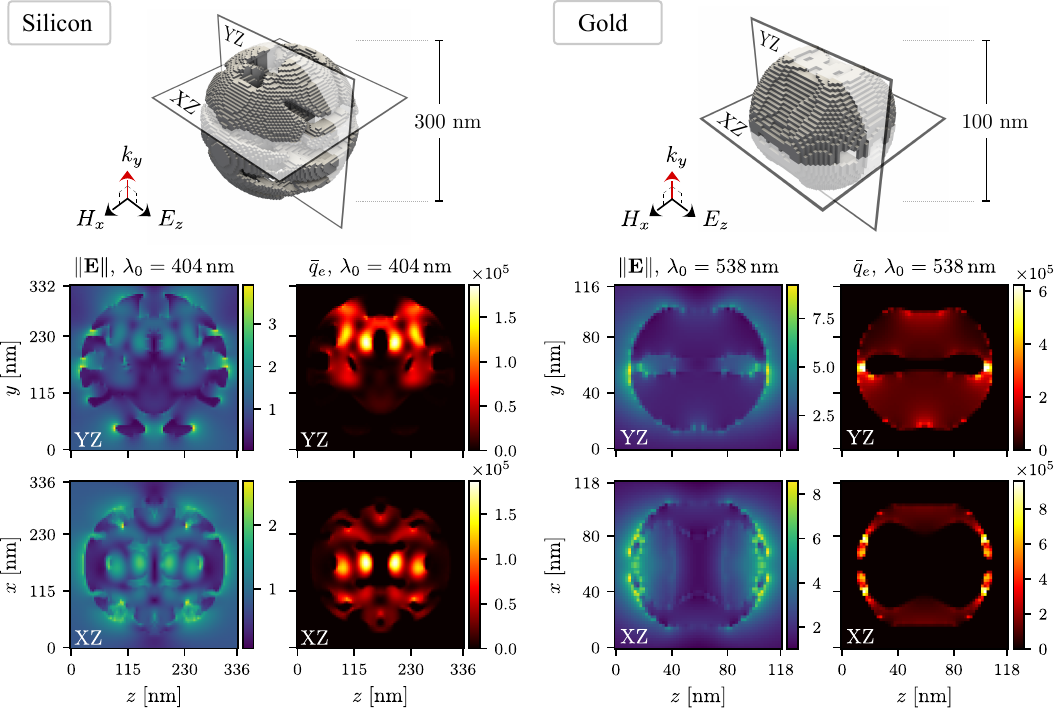}}
\caption{Cross-sections through the optimized designs made of Silicon and Gold, showing the amplitude of the electric field $\|\mathbf{E}\|$ and the time-averaged electric power dissipation density $\bar{q}_{e}$ for the wavelength at which the corresponding design has its absorption maximum. The amplitude of the electric field is normalized to that of the incident field. The unit of the dissipation density is given in $\mathrm{W}/\mathrm{m}^{3}$. The planes were selected based on where the dissipation density shows particularly strong local hotspots; see Figs.~\ref{fig:performance} (c.1)-(c.2).}
\label{fig:slices}
\end{figure*}

We optimized a spherical Silicon and Gold nanoparticle surrounded by air ($\varepsilon^{(1)}= 1$) for having an enhanced absorption efficiency $Q_{\text{abs}}$ in the spectral range 300-700 nm. A summary of the parameters used for the FDTD simulations and topology optimization can be found at the end of this section (Sec.~\ref{sec:Parameters}). The projection parameters and artificial damping term $\gamma$ for Gold were selected based on the preliminary considerations from Sec.~\ref{sec:dampingterm}. Both materials have been modelled by the CCPR parameters listed in Ref.~\cite{Gedeon2023} and the corresponding fits are shown in Fig.~\ref{fig:Epsilon}. Two poles were used for the fit of Silicon, and three poles for Gold. The systems were excited by a $z$-polarized plane wave, covering the frequencies of interest. We restricted the designs to a spherical volume with diameter $d$, which was chosen to be $d=300$ nm for Silicon and $d=100$ nm for Gold. Here, we compromised between highlighting material-dependent features evolving due to different physical effects appearing in dielectric and plasmonic structures, while ensuring numerical accuracy and a reasonable simulation time and memory usage when storing the fields during the optimization. A sufficient numerical accuracy was validated by simulating the response of a (binary) sphere with the same setup employed in the design optimizations, followed by a comparison of the computed dissipation via Eq.~(\ref{Eq:FreqDissipation}) against the analytical results from Mie theory \cite{Mie1908, MieCode}.\par The convergence of both structures is illustrated in \mbox{Figs.~\ref{fig:Convergence} (a)-(b)}. The figures show the normalized objective function with respect to that of a sphere sharing the same diameter, and the measure of non-discreteness per iteration step. In both cases, we started with a filtered and projected sphere as the initial geometry. We let the Silicon and Gold design evolve for 50 and 100 iterations, respectively. A stable convergence behavior for both materials can be observed, and they tend to evolve to a binary structure. In the case of Gold, we recognize a jump in performance after the final threshold - and from iteration 80 onwards, the design shows no significant improvement in terms of performance and its tendency towards a binary structure. In fact, we observed that the Gold structure retains localized gray areas on the inner surface of the centrally formed cavity. Increasing the projection parameter $\beta$ could not remedy this. As previously described in Sec.~\ref{sec:dampingterm}, we can attribute this to the optimization driven by propagating, spatially confined but deeply penetrating surface plasmons, and after the threshold, the localized surface plasmon resonances come into full effect.\par 
The results of the optimization are presented in \mbox{Figs.~\ref{fig:performance} and \ref{fig:slices}}, illustrating the topologically designed structures, their performance compared to a sphere with the same diameter $d$, and the electric field and local dissipation distribution for the wavelengths of maximum absorption. The absorption efficiencies $Q_{\text{abs}}$ from Eq.~(\ref{Eq:Qabs_eq}) have been computed by integration of Eq.~(\ref{Eq:FreqDissipation}) over the design's volume together with the electric fields from the FDTD simulation (via discrete Fourier transformation), considering an intensity of the incident wave of 1 $\;\mathrm{W/m^{2}}$. The geometrical cross section was determined by the given spherical restriction of the designs.\par The spherical structure made of Silicon is characterized by a series of interlocked material-free segments that enveloped at its surface, \mbox{Fig.~\ref{fig:performance} (a.1)}. As we can see from the performance plot in \mbox{Fig.~\ref{fig:performance} (b.1)}, the design demonstrates a significant increase in absorption compared to that of a sphere in the range \mbox{300-550 nm}. We note that both absorption peaks of a sphere, corresponding to the electric quadrupole and magnetic octupole modes in the range \mbox{630-670 nm} \cite{ITMO}, are suppressed in the case of the topologically optimized structure. They are correlated to resonances characterized by a high quality factor, which are associated with a long-lasting duration of the oscillations (Appendix~\ref{app:Silicon_compr}). As a consequence, it will effectively contribute to a higher absorption of energy over time. We can explain the evolution of the design primarily favoring the increase in absorption for \mbox{$\lambda<$  550 nm} by recalling the following: The maximization of the dissipation in the time domain is equivalent to maximizing the integral value over the entire spectrum of frequencies contained in the time-dependent source of excitation. The evolving design is therefore not restricted to keep or produce sharp resonances at specific wavelengths, as long as it yields a high absorption over a broadband range in return. Furthermore, the imaginary part of the permittivity of Silicon gets larger towards the ultraviolet regime. Thus, the optimizer seems to exhibit a propensity to leverage this characteristic for the maximization of the objective function. Reproducing single sharp resonances associated with higher absorption is limited due to the nature of the computational approach itself. In this case, the limitation is imposed by the simulation time. That can be understood by noticing the discrepancy between the numerically obtained peak at \mbox{$\lambda\approx$ 645 nm} of the sphere and that predicted by Mie theory, \mbox{Fig.~\ref{fig:performance} (b.1)}. We recall that an increase in simulation time yields a more accurate result but also increases the time per iteration and memory consumption due to the storage of fields (in space and time) during the optimization. In order to better understand the optical characteristics of the optimized Silicon design, we analyzed the resonance spectrum of the design compared to a sphere in Appendix~\ref{app:Silicon_compr}. We observed that our design's topology supports the formation of a cascade of high-quality-factor resonances in the wavelength range $<$ 550 nm, which get effectively absorbed.\newline 
Fig.~\ref{fig:performance} (c.1) gives an insight into the spatial profile of the electric field and dc component of the dissipation, Eq.~(\ref{Eq:FreqDissipation}), for the wavelength \mbox{$\lambda_{0}=$ 404 nm}. Here, a comparison with a sphere is provided in addition. The electric field's amplitude is confined in the material-free notches on the surface of the designed structure. By examining the spatial dissipation distribution within the structure itself, we observe numerous diffusely distributed hotspots that collectively enhance the absorption of the electric field's energy. A clearer picture of this phenomenon is provided by Fig.~\ref{fig:slices}, which shows cross-sections depicting the localization of particularly strong hotspots.\par
The Gold design shown in Fig.~\ref{fig:performance} (a.2) appears as a more compact structure. Three distinct holes on the surface on either side with respect to the $z$-axis form the entrance to an inner cavity. From the performance plot in Fig.~\ref{fig:performance} (b.2) we observe a $\approx$ 40\% improvement at the absorption peak corresponding to the electric dipole mode resonance, which is slightly shifted towards larger wavelengths. The design yields an amplification of $Q_{\text{abs}}$ across almost the entire spectrum compared to that of the sphere. The spatial electric field and the dissipation profile for \mbox{$\lambda_{0}=$ 538 nm} in Fig.~\ref{fig:performance} (c.2) and their cross-sections in Fig.~\ref{fig:performance} (c.2) provide insight into the topological formation of the cavity with regard to the plasmonic effects. The electric field is mainly confined at the surface, where thin walls separate the outside from the inner cavity. In these walls, the local dissipation has its maximum. This is understandable if we recall the physical nature of plasmons. The topology enables a balance between their formation and longevity along the surface, and their penetration into the material itself, which is limited by the skin depth inside the metal.\par
We conclude that the optimization worked successfully for both material types, and a broadband performance could be achieved in each case. We note that the choice of the artificial damping parameter $\gamma$ for Gold was crucial, as a too small or large value impairs the convergence or leads to the design remaining gray. Despite the remarkable broadband performance of the optimized Gold structure, we observed that the corresponding density retained thin gray areas along the surface of the inner cavity. Neither an increase in iteration steps nor modulation of the projection parameter $\beta$ could circumvent it. In the case of Silicon, we performed additional optimizations with different imposed diameters of the spherical restriction (Appendix~\ref{app:Silicon_compr}). We found that as the diameter was increased, the optimized designs yield an improvement in performance towards larger wavelengths. This is plausible as more space for the design to develop is provided. Thus, the evolving structure can utilise the dissipative effects related to resonances at longer wavelengths.

\subsection{Optimization and simulation parameters}\label{sec:Parameters}
\begin{table}[H]
\begin{figure}[H]
    \centering
    \includegraphics[width=0.95\linewidth]{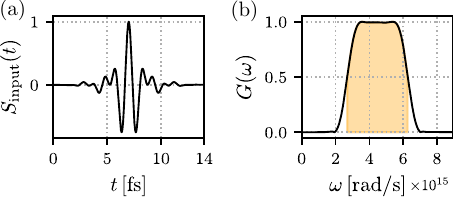}
    \caption{(a) Time and (b) frequency domain plots of the excitation source's amplitude. The area in yellow marks the frequency range of interest, bounded by the values $\omega_{1} = 6.28 \times 10^{15}\, \mathrm{rad/s}$ and $\omega_{2} = 2.7 \times 10^{15}\, \mathrm{rad/s}$ of the half bandwidth, which correspond to the wavelengths of 300 nm and 700 nm, respectively.}
    \label{fig:Signal}
\end{figure}
\caption{TopOpt and FDTD configuration settings used for designing the spherical nanostructures presented in Sec.~\ref{sec:Results}}
\label{table}
\setlength{\tabcolsep}{3pt}
\begin{tabular}{|p{80pt}|p{80pt}|p{80pt}|}
\hline
Parameter& 
Silicon & 
Gold \\
\hline
Simulation domain& 
\mbox{$125\times125\times125$ cells} & 
\mbox{$100\times100\times100$ cells}\\
Design domain $\Omega$& 
\mbox{$86\times85\times86$ cells} & 
\mbox{$61\times60\times61$ cells}\\
Observation domain $\Omega_{\mathrm{o}}$& 
\mbox{$86\times85\times86$ cells} & 
\mbox{$61\times60\times61$ cells}\\
CPML  & 
25 cells & 
20 cells \\
d$x$, d$y$, d$z$ & 
$4$ nm & 
$2$ nm\\
d$t$ & 
6.67 as & 
3.34 as\\
Filter radius $R$ & 
12 nm & 
10 nm\\
Time steps per sim. $M$& 
22000  & 
12000 \\
$\eta$& 
0.5  & 
0.55 \\
$\beta_{0}$& 
10  & 
7 \\
$\beta_{\mathrm{max}}$& 
20  & 
42 \\
$\gamma$& 
0  & 
$2\times 10^{5}$ \\
TopOpt iterations& 
50  & 
100 \\
\hline
\end{tabular}
\label{tab1}
\end{table}
Table \ref{tab1} shows the configuration of both TopOpt and FDTD parameters used for the optimizations presented in the previous section.
The simulation domain was surrounded by convolutional perfectly matched layers (CPML) to mimic open boundaries by minimizing the reflection from the edges of the simulation grid. \mbox{Design domain $\Omega$} and observation region $\Omega_{\mathrm{o}}$ were placed in the center. They were both cubic and shared the same dimensions, i.e., $\Omega \equiv \Omega_{\mathrm{o}}$. We chose a homogeneous sphere with $\rho\equiv\eta$ as our initial geometry, which was subsequently filtered and projected. We started  with an initial projection parameter $\beta_{0}$ and let the value increase gradually during the optimization until it reached a maximum value of $\beta_{\mathrm{max}}$. During the forward simulations, the system was excited by a \mbox{$z$-polarized} plane wave carrying a truncated sinc signal covering the frequencies of interest, Fig.~\ref{fig:Signal}. The number of time steps $M$ per forward and adjoint simulation (Algorithm 1) was determined by whether the absorption spectrum of a sphere matches that of the Mie theory with sufficient accuracy (see Fig.~\ref{fig:performance} (b.1)-(b.2)). In addition, it was ensured that there was no significant change in the objective function during the optimization after this time period (due to the decay of the fields). We let the Silicon and Gold design evolve for 50 and 100 iterations, respectively. After each iteration, the density was updated using the Method of Moving Asymptotes (MMA) \cite{Svanberg1987} and subsequently stenciled out via an indicator function, $\rho \cdot \mathbf{1}_{\mathcal{B}_{d/2}(\bm{r}_0)}$, to restrict the evolving design to a spherical volume with a fixed diameter $d$. Here, $\bm{r}_0$ denotes the center point of the domain $\Omega$, and $\mathcal{B}_{d/2}(\bm{r}_0)$ describes a sphere with radius $d/2$ around $\bm{r}_0$. The diameter was chosen to be $d=300$ nm for Silicon and $d=100$ nm for Gold.\newline
The optimizations were performed with our parallelized in-house FDTD software with TopOpt implementation \cite{Gedeon2023} on the supercomputer HLRN-IV-System Emmy in G\"ottingen, Germany, provided by the North German Supercomputing Alliance as part of the National High Performance Computing (NHR) infrastructure. The simulation domain has been decomposed into multiple chunks \cite{parallelism} in order to accelerate the optimization and to ensure an efficient memory distribution when storing the electric and auxiliary fields in space and time.
The optimization of the Silicon structure presented in Sec.~\ref{sec:Results} took $\approx$ 11 hours on 2304 CPUs, while the optimization for the Gold structure took $\approx$ 10 hours using 1152 \mbox{CPUs (Cascade 9242)}. 

\section{Conclusion}\label{sec:Conclusion}
We introduced a novel time-domain gradient-based topology optimization method to locally optimize the electric power dissipation for dispersive materials over a broadband range of frequencies. The method is formulated based on the CCPR model to accurately describe the material's dispersion within any desired spectral range. By using the auxiliary differential equation (ADE) method, we derived the expression for the instantaneous electric power dissipation based on the CCPR model, and formulated an adjoint scheme to obtain the gradient information of the time-averaged dissipation defined as our objective function. The discretization and integration of the adjoint system into the finite-difference time-domain (FDTD) framework was explained in detail. Our method was successfully demonstrated on the optimization of spherical nanoparticles made of Silicon and Gold for an enhanced absorption efficiency in the visible and ultraviolet regimes.\par  
The adjoint scheme presented in Sec.~\ref{sec:AdjointMethod} is kept in a general form and is thus not restricted to the specific setup used in our study case. It further allows for other configurations, such as imposing periodic boundary conditions to enable the design of broadband absorbing metasurfaces, considering both background and design materials as dispersive to optimize hybrid nanostructures, or maximizing dissipation in a specific subvolume $\Omega_{\mathrm{o}}\subset\Omega$ to generate localized heat sources (e.g., for thermal ablation). 
We note that despite the success, the optimization comes with challenges, as it was shown by the theoretical consideration of the plasmonic effects of Gold during optimization and its convergence behavior in practice. We can attribute this to the density-based approach in the context of maximizing dissipation per se, which requires appropriate tuning with respect to a material parameter for intermediate densities. We assume that a refinement of the material interpolation can lead to even better results. It is also possible that the adaptation and implementation of the latest achievements in shape optimization could be promising for tackling this problem \cite{Keenan}, which completely dispenses with the concept of material interpolation.
\par Our contribution holds great potential for tackling inverse design problems, where dispersive materials are included, and a reduced or enhanced absorption over a broadband range is desired. Areas of application include solar energy harvesting, by maximizing light absorption in photovoltaic solar cells or improving thermal radiation absorption in thermophotovoltaics; broadband metasurface absorbers, selective plasmonic sensors and absorptive filters; anti-reflective or high-absorption coatings; devices with thermal tuning exploiting thermal effects to vary the refractive index; or designing low-loss devices like waveguides, couplers, and antennas with improved efficiency.
\appendices
\section{Time average of the electric power dissipation density for the time-harmonic case}\label{app:Dissip_freq_domain}
In the time-harmonic case, the instantaneous electric power dissipation density in Eq.~(\ref{Eq:InstantanousDissipation}) can be split into a dc and a sinusoidal ac term \cite{Shin:12}. To measure the broadband performance of an optimized device in the steady state regarding the time-average of that dissipation density, we extract its dc component here. \newline
Consider a time harmonic field $\mathbf{E}(t):= \Re\{\mathbf{\hat{E}}_0e^{j\omega_{0} t}\}$ oscillating with a frequency $\omega_{0}$, where $\hat{\mathbf{E}}_{0}$ itself is frequency independent. The Fourier transform yields
\begin{equation}
\hat{\mathbf{E}}(\omega)=\frac{1}{2}\left(\hat{\mathbf{E}}_{0}\,\delta\left(\omega_0-\omega\right)+\hat{\mathbf{E}}_{0}^{*} \,\delta\left(\omega_0+\omega\right)\right).
\end{equation}
The auxiliary field components in the frequency domain based on the CCPR model in Eq.~(\ref{Eq:eps_CCPR}) are related to the electric field components via \cite{material}
\begin{equation}
\hat{\mathbf{Q}}_p(\omega)=\frac{\varepsilon_0 c_p}{j \omega-a_p} \hat{\mathbf{E}}(\omega).
\end{equation}
Thus, based on the time-harmonic field, the auxiliary field in time domain and its derivative read
\begin{align}
\mathbf{Q}_p(t)=&\int_{-\infty}^{+\infty} \hat{\mathbf{Q}}_p(\omega) e^{j \omega t} d \omega \\
=&\,\frac{1}{2}\left(\frac{\varepsilon_0 c_p}{j \omega_0-a_p} \hat{\mathbf{E}}_{0} e^{j \omega_0 t}+\frac{\varepsilon_0 c_p}{-j \omega_0-a_p} \hat{\mathbf{E}}_{0}^{*} e^{-j \omega_0 t}\right), \notag \\[4pt]
\partial_{t} \mathbf{Q}_p(t)=&\,\frac{1}{2}\left(\frac{j \omega_0 \varepsilon_0 c_p}{j \omega_0-a_p} \hat{\mathbf{E}}_{0} e^{j \omega_0 t}+\frac{-j \omega_0 \varepsilon_0 c_p}{-j \omega_0-a_p} \hat{\mathbf{E}}_{0}^{*} e^{-j \omega_0 t}\right).
\end{align}
Substituting $\partial_{t} \mathbf{Q}_p$ into our formula for the instantaneous electric power dissipation density, 
\begin{equation}\label{Eq:InstantanousDissipation2}
q_e(t) = \sigma \mathbf{E}^{2} + 2\sum_{p=1}^P \Re\left\{\frac{(\partial_{t}\mathbf{Q}_p)^2}{\varepsilon_{0}c_p}\right\},
\end{equation}
 yields an expression consisting of both the ac and dc terms. Since we consider the time-average of the dissipation as our objective, we are only interested in its dc part. Using the identity
\begin{equation}
\Re\{\mathbf{W}\}^2=\frac{1}{4}\left(\mathbf{W}+\mathbf{W}^*\right)^2=\frac{1}{2}|\mathbf{W}|^2+\frac{1}{2} \Re\left\{\mathbf{W}^2\right\},
\end{equation}
that holds for any complex vector field $\mathbf{W}$, and neglecting all the terms containing the factors $e^{\pm 2j\omega t}$, leaves the dc component, 
\begin{align}\label{Eq:Dissipation_Freq}
&\bar{q}_{e}(\omega_{0})=\left(\frac{1}{2}\sigma 
+ \sum_{p=1}^{P}\Re\left\{\frac{\omega_{0}^{2} \varepsilon_{0} c_p}{(j\omega_{0} - a_p)(-j\omega_{0} - a_p)}\right\}\right)|\hat{\mathbf{E}}_{0}|^{2}.
\end{align}
We can further express this result in terms of the imaginary part of the permittivity $\varepsilon$. By noticing that
\begin{align}
&\,\frac{1}{2j}\left( \frac{c_p}{j\omega_{0} - a_p}  - \frac{c_p}{-j\omega_{0} - a_p}\right) + \mathrm{c.c.}  \\
=& \,\frac{1}{2j}\left\{\left( \frac{c_p}{j\omega_{0} - a_p}  + \frac{c_p^{*}}{\phantom{-}j\omega_{0} - a_p^{*}}\right) - \mathrm{c.c.}\right\},
\end{align}
and using the identities \mbox{$\Re\{z\} = \frac{1}{2}(z + z^{*})$} and \mbox{$\Im\{z\} = \frac{1}{2j}(z - z^{*})$} for complex numbers $z$, the prefactor in brackets in Eq.~(\ref{Eq:Dissipation_Freq}) can be written as
\begin{align}\label{Eq:Landau_equivalenz}
&\;\frac{\sigma}{2} 
+ \sum_{p=1}^{P}\Re\left\{\frac{\omega_{0}^{2} \varepsilon_{0} c_p}{(j\omega_{0} - a_p)(-j\omega_{0} - a_p)}\right\} \\
=&\;\frac{\sigma}{2}  -\varepsilon_{0}\omega_{0} \sum_{p=1}^{P} \Re\left\{\frac{1}{2j}\left( \frac{c_p}{j\omega_{0} - a_p}  - \frac{c_p}{-j\omega_{0} - a_p} \right)\right\} \\
=&-\frac{1}{2}\varepsilon_{0}\,\omega_{0} \,\Im\left\{ \varepsilon(\omega_{0}) \right\}.
\end{align}
From that, we observe that Eq.~(\ref{Eq:Dissipation_Freq}) is equivalent to the model-independent expression derived by Landau and Lifschitz in Ref.~\cite{landau1995electrodynamics},
\begin{equation}
\bar{q}_{e}(\omega_{0})= \frac{1}{2}\varepsilon_{0}\,\omega_{0} \,\Im\left\{ \varepsilon(\omega_{0}) \right\}|\hat{\mathbf{E}}_{0}|^{2},    
\end{equation}
considering the convention \mbox{$\varepsilon= \varepsilon^{\prime} - j\varepsilon^{\prime\prime}$}.
\section{Discretization of the adjoint equations}\label{app:discretization}
Here, we present the update equations for the adjoint fields in time, considering the time-averaged dissipation in Eq.~(\ref{Eq:ObjectiveDissipation}) to be the objective function. The update scheme for the forward equations is identical, except for the time shift of $\Delta t / 2$ and the absence of the adjoint source terms. In that case, the equations are reduced to those presented in Ref.~\cite{material}. We again consider isotropic media for both background and design material, and denote these materials with the indices $i=1,2$, respectively. The material parameters $\varepsilon_{\infty}$, $\sigma$ appearing in the following equations are assumed to be material interpolated by the density $\rho$, and the prefactor $\kappa^{(i)}$ is defined as  $\kappa^{(1)}(\rho):=1-\rho$ and $\kappa^{(2)}(\rho):= \rho$. The extension to anisotropic media can be found in the Supplementary Material.\par 
We discretize Eq.~(\ref{Eq:Adjoint}a) and all Eqs.~(\ref{Eq:Adjoint}b) at time steps $m\Delta t$. Since the electric and auxiliary fields of the forward system are considered to be discretized at time steps $m\Delta t$ too (Fig.~\ref{fig:Discretization}), we can discretize the adjoint source term $\mathbf{S}_{E}$ depending on the forward electric field as
\begin{align}
\mathbf{S}^{m}_{E}= 2T^{-1}\sigma \overleftarrow{\mathbf{E}}\left[m\Delta t\right] = 2T^{-1}\sigma \mathbf{E}^{M-m},
\end{align}
and each adjoint source term $\mathbf{S}_{\partial{\tau}Q_{p}}^{m, (i)}$ depending on the time-derivative of the corresponding forward auxiliary field as
\begin{align}
\mathbf{S}_{\partial{\tau}Q_{p}}^{m, (i)}=&\;2T^{-1}\partial_{\tau}\overleftarrow{\mathbf{Q}}_{p}^{(i)}\left[m\Delta t\right] \\
=&\;2T^{-1}\;\frac{\left(\mathbf{Q}_{p}^{M-m-1,(i)}-\mathbf{Q}_{p}^{M-m+1,(i)}\right)}{2\Delta t},
\end{align}
where the centered difference approximation was employed. The update equation for each adjoint auxiliary field is then  
\begin{align}\label{Eq:Q_auxiliary_update}
\tilde{\mathbf{Q}}_{p}^{m+ \frac{1}{2}, (i)}=&\;\frac{2+a_{p}^{(i)} \Delta t}{2-a_{p}^{(i)} \Delta t} \tilde{\mathbf{Q}}_{p}^{m - \frac{1}{2}, (i)} \\
+&\;\frac{\varepsilon_0 c_{p}^{(i)} \Delta t}{2-a_{p}^{(i)} \Delta t}\left(\tilde{\mathbf{E}}^{m+\frac{1}{2}} + \tilde{\mathbf{E}}^{m-\frac{1}{2}}\right) \notag\\ 
+ &\;\frac{2\Delta t}{2-a_{p}^{(i)} \Delta t}\mathbf{S}_{\partial{\tau}\mathbf{Q}_{p}}^{m, (i)}. \notag
\end{align}
Using these expressions, the update equation of the electric field reads
\begin{equation}
\tilde{\mathbf{E}}^{m+\frac{1}{2}}=\alpha^{-1}\mathbf{g}^{m-\frac{1}{2}},
\end{equation}
where $\alpha$ is a density-dependent constant defined as
\begin{align}
\alpha:= \frac{\varepsilon_{0}\varepsilon_{\infty}}{\Delta t}  + \frac{\sigma}{2} +\frac{2}{\Delta t}\sum_{i =1}^{2}\sum_{p=1}^{P^{(i)}}\kappa^{(i)}\Re\left\{\frac{\varepsilon_0 c_{p}^{(i)} \Delta t}{2-a_{p}^{(i)} \Delta t}\right\},
\end{align}
and $\mathbf{g}^{m-\frac{1}{2}}$ changes at every time step according to
\begin{align}\label{Eq:g_update}
\mathbf{g}^{m-\frac{1}{2}}=&\;(\nabla \times \tilde{\mathbf{H}})^{m} + \left(\frac{2\varepsilon_{0}\varepsilon_{\infty}}{\Delta t} - \alpha\right)\tilde{\mathbf{E}}^{m-\frac{1}{2}}\\
-&\;\frac{2}{\Delta t}\sum_{i=1}^{2} \sum_{p=1}^{P^{(i)}} \kappa^{(i)}\Re\left\{\frac{2a_{p}^{(i)} \Delta t}{2-a_{p}^{(i)} \Delta t}\tilde{\mathbf{Q}}_{p}^{m-\frac{1}{2},(i)}\right\} \notag\\
-&\;\frac{2}{\Delta t}\sum_{i=1}^{2}\sum_{p=1}^{P^{(i)}} \kappa^{(i)}\Re\left\{\frac{2\Delta t}{2-a_{p}^{(i)} \Delta t}\mathbf{S}_{\partial{\tau}\mathbf{Q}_{p}}^{m, (i)} \right\} \notag\\[6pt]
+&\; \mathbf{S}^{m}_{E}. \notag
\end{align}

\section{Mie analysis of the Silicon TopOpt design}\label{app:Silicon_compr}
\begin{figure}[H]
\centerline{\includegraphics[width=0.9\linewidth]{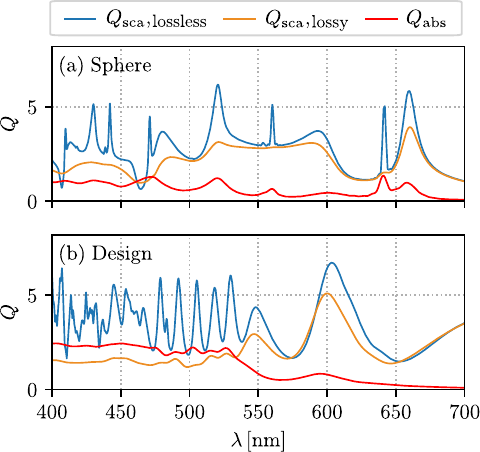}}
\caption{Scattering efficiency \(Q_{\mathrm{sca},\,\mathrm{lossy}}\) (orange) and absorption efficiency \(Q_{\mathrm{abs}}\) (red) of both Silicon (a) sphere and (b) TopOpt design presented in Sec.~(\ref{sec:Results}) for the wavelength range \mbox{400-700 nm}. To identify the resonances, the scattering efficiency \(Q_{\mathrm{sca},\,\mathrm{lossless}}\) (blue) is also shown here for the undamped case, i.e., setting \(\Im\{\varepsilon\}=0\).}
\label{fig:Qsca}
\end{figure}

For a better understanding of the optical properties of the optimized Silicon design from Sec.~\ref{sec:Results}, we examined the resonances of the optimized structure compared to the sphere (Fig.~\ref{fig:Qsca}). To identify the resonances of both structures, we illustrate not only their scattering efficiency \(Q_{\mathrm{sca},\,\mathrm{lossy}}\) and absorption efficiency \(Q_{\mathrm{abs}}\), but also the scattering efficiency \(Q_{\mathrm{sca},\,\mathrm{lossless}}\) for the artificial undamped case, i.e., setting \(\Im\{\varepsilon\}=0\). Here, the spectral range is limited to \mbox{400-700 nm}, in which it was observed that the optimized design tends to outperform the sphere in terms of absorption efficiency towards smaller wavelengths (see Fig.~\ref{fig:performance} (b.1)). From the $Q_{\mathrm{sca},\,\mathrm{lossless}}$ plots, it can be observed that our design exhibits a denser distribution of resonances in the range of \mbox{400-550 nm} compared to the sphere. The sharp peaks (high-$Q_{f}$ resonances) are associated with higher quality factors $Q_{f}$, which correspond to a longer lifetime $\tau = 2Q_{f}(\omega)/\omega$ of the resonances~\cite{saleh2019fundamentals}.
As a consequence, more energy can be absorbed in this period of time if we include the damping, which effectively contributes to increasing our objective function in Eq.~(\ref{Eq:ObjectiveDissipation}). Considering the damping of Silicon, we see a stronger reduction of scattering and an increase of absorption compared to the sphere, which indicates an efficient absorption of the resonances. Figure~\ref{fig:Qsca} also shows why the optimized design exhibits relatively weak absorption in the 630-700 nm range. High-$Q_f$ resonances, as in the sphere case, are completely absent here.\par
\begin{figure}[H]
\centerline{\includegraphics[width=.9\linewidth]{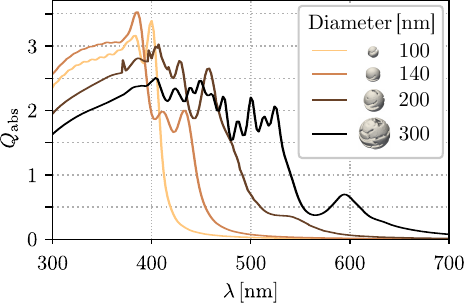}}
\caption{Absorption efficiencies \(Q_{\text{abs}}\) of topology optimized nanostructures made of Silicon for different diameters, by which the designs were restricted in their spatial extension. The absorption efficiency corresponding to a diameter of 300 nm corresponds to the design presented in Sec.~\ref{sec:Results}.}
\label{fig:Silicon_diffRadii}
\end{figure}

 
From these observations, we can conclude that the topological shaping of the design leads to the formation of a cascade of high-$Q_f$ resonances, which in turn enhance the absorption. It is therefore reasonable to analyze the performance of the optimized designs in relation to the design volume. Restricting the spherical design to a smaller diameter reduces the degrees of freedom available for the development of the topology, which is needed for the emergence of multiple high-Q resonances across the entire spectral range of \mbox{300-700 nm}. To verify that, we performed topology optimizations of spherical Silicon particles for different diameters. To ensure an optimal trade-off between numerical precision, computational time, and memory consumption, we restricted our optimizations to diameters up to 300 nm. The results are illustrated in Fig.~\ref{fig:Silicon_diffRadii}. We can clearly see that an increase in the diameter leads to an improvement in broadband performance. The absorption curve corresponding to a diameter of \mbox{300 nm} is the performance of the design presented in Sec.~\ref{sec:Results}. We can expect that increasing the radius further will result in even better broadband absorption, as the increased spatial extent and topological complexity allow the appearance of resonances at longer wavelengths too. 

\bibliographystyle{IEEEtran.bst}
\bibliography{bibliography.bib}

\end{document}


\title{Supplementary Material for \\
``Time-Domain Topology Optimization of Power Dissipation in Dispersive Dielectric and Plasmonic Nanostructures''}
\author{Johannes Gedeon, Izzatjon Allayarov, Antonio Cal{\`a} Lesina, and Emadeldeen Hassan
}

\maketitle

This supplementary material provides a detailed derivation of the adjoint scheme for density-based topology optimization in the time domain of anisotropic, dispersive materials for two important classes of objectives:
\begin{enumerate}
\item An objective function that can be represented as a functional $F[\mathbf{E}]$ dependent solely on the electric field (Section~\ref{sec:Adjoint_method_E_dep}). This includes, for example, the scenarios of maximizing the electric field energy $\varepsilon \mathbf{E}^{2}$ in non-dispersive media or the electric power dissipation $\sigma \mathbf{E}^{2}$, where the losses are described by the static conductivity $\sigma$. In these cases, the material distribution within the observation volume $\Omega_{o}$ is considered fixed, meaning the density remains constant throughout the optimization iterations.

\item The electric power dissipation within a volume $\Omega_{o}$ containing lossy, dispersive material (Section~\ref{sec:adjoint_powerdissipation}). In that case, a simple description based only on the electric field is insufficient. This requires an extension of the adjoint scheme provided in Section~\ref{sec:Adjoint_method_E_dep}, which also encompasses the objective in case 1 related to dissipation as a trivial special case.
\end{enumerate}
Both classes of objectives cover the fundamental physical quantities of electric field energy and power dissipation, which are present in Poynting's theorem. The treatment of dispersion is incorporated into Maxwell's equations using the Auxiliary Differential Equation (ADE) method, based on the complex-conjugate pole-residue (CCPR) model of the permittivity. The implementation of the adjoint scheme within the FDTD framework is described in Section~\ref{sec:discretization}.\newline\newline
All the following derivations are built on the description of the relative permittivity $\varepsilon$ by the CCPR model with $\exp^{\,j\omega}$ time-dependency \cite{material},
\begin{equation}\label{Eq:CCPR_model}
\varepsilon_{\alpha \beta}(\omega)=\varepsilon_{\infty, \alpha \beta}+\frac{\sigma_{\alpha \beta}}{j \omega \varepsilon_0}+\sum_{p=1}^{P_{\alpha \beta}}\left(\frac{c_{p, \alpha \beta}}{j \omega-a_{p, \alpha \beta}}+\frac{c_{p, \alpha \beta}^*}{j \omega-a_{p, \alpha \beta}^*}\right),
\end{equation}
where $\varepsilon_{\infty, \alpha \beta}$ is the relative permittivity at infinite frequency, and $\sigma_{\alpha \beta}$ is the static conductivity. The indices $\alpha$ and $\beta$ denote the $x$, $y$ and $z$ component and $*$ represents the complex conjugation. We denote the background material with index $i=1$, and the design material with index $i=2$. For a given density value $\rho \in [0, 1]$ we apply a linear mixing of the parameters $\varepsilon_{\infty,\alpha \beta}$ and $\sigma_{\alpha \beta}$, and complex pole pairs,
\begin{subequations}\label{Eq:InterpolatedModel}
  \begin{empheq}[]{align}
&\varepsilon_{\infty, \alpha \beta}(\rho) := \sum_{i \in \{1,2\}}\kappa^{(i)}(\rho)\, \varepsilon_{\infty, \alpha \beta}^{(i)},\\
&\sigma_{\alpha \beta}(\rho) := \sum_{i \in \{1,2\}}\kappa^{(i)}(\rho) \,\sigma_{\alpha \beta}^{(i)}\,\, + \rho\,(1-\rho)\,\gamma,\\
&{\textstyle \sum_{{\alpha \beta}}}(\omega, \rho) := \nonumber \\
 &\quad\sum_{i \in \{1,2\}}\kappa^{(i)}(\rho)\sum_{p=1}^{P_{\alpha \beta}^{(i)}}\left(\frac{c_{p, \alpha \beta}^{(i)}}{j \omega-a_{p, \alpha \beta}^{(i)}}+\frac{c_{p, \alpha \beta}^{(i)*}}{j \omega-a_{p, \alpha \beta}^{(i)*}}\right),
\end{empheq}
\end{subequations}
where $\kappa^{(1)}(\rho): = (1-\rho)$ and $\kappa^{(2)}(\rho): = \rho$, and $\gamma$ is an artificial damping parameter. According to these equations, the interpolated relative permittivity can be written as
\begin{equation}\label{Eq:CCPR_model_mixed}
\varepsilon_{\alpha \beta}(\omega, \rho) = \varepsilon_{\infty, \alpha \beta}(\rho) + \frac{\sigma_{\alpha \beta}(\rho)}{j \omega \varepsilon_0} + {\textstyle \sum_{{\alpha \beta}}}(\omega, \rho).
\end{equation}

\section{Sensitivity analysis for an objective $F[\mathbf{E}]$}\label{sec:Adjoint_method_E_dep} 
We assume that the objective function we aim to maximize, is a functional of the electric field only, such that the optimization problem can be formulated as follows:
\begin{equation}\label{Eq:objective_nondispersive_fixedDens}
\begin{aligned}
& \max _{\rho} F[\mathbf{E}] \\
& \text { s.t. Maxwell's equations,}
\end{aligned}
\end{equation}
where boundary conditions and manufacturability constraints can be included. The functional derivate of the objective $F[\mathbf{E}]$ with respect to the density is
\begin{equation}\label{Eq:ObjectiveDerivative}
\frac{\delta F[\mathbf{E}]}{\delta \rho}=\frac{\delta F[\mathbf{E}]}{\delta \mathbf{E}} \cdot \frac{\mathrm{d} \mathbf{E}}{\mathrm{d} \rho},
\end{equation}
where the first multiplicator on the right hand side denotes the functional derivate of the objective function with respect to the electric field. We denote the derivative of any local function with respect to the density by ``$\mathrm{d}_{\rho}$'' in the following.\par 
We assume a diagonal permittivity tensor and non-magnetic materials in the following. We refrain from marking the density dependency for the sake of readability, i.e., $\varepsilon_{\infty} := \varepsilon_{\infty}(\rho)$,  $\sigma := \sigma(\rho)$, and $\kappa^{(i)}:=\kappa^{(i)}(\rho)$, but remark that the dependency is essential in the following derivations.
We denote the design domain with $\Omega$, and the time interval as $I=[0,T]$, and consider an excitation of the \textit{forward} system by a pulse injected at $\partial \Omega$ at $t=0$, and vanishing fields for $t\in \partial I$. The full system of Maxwell equations for each spatial component $k\in\{1,2,3\}$ and $\forall (\mathbf{r},t) \in \Omega \times I$ of the forward system reads \cite{material}:
\begin{subequations}\label{Eq:Maxwells2}
  \begin{empheq}[]{align}
-(\nabla \times \mathbf{H})_{k} + \varepsilon_{0}\varepsilon_{\infty, k}\partial_{t}E_{k} + \sigma_{k} E_{k} +\qquad &   \nonumber \\  + 2\sum_{i \in \{1,2\}}\sum_{p=1}^{P_{k}^{(i)}}\kappa^{(i)}\Re\left\{\partial_{t}Q_{p, k}^{(i)}\right\} &= 0,\\[-3pt]
\text{For } i =1,2 \text{ and } \forall p \in \{1, \ldots, P_k^{(i)}\}: & \nonumber \\ \partial_{t}Q_{p, k}^{(i)}-a_{p, k}^{(i)}Q_{p, k}^{(i)}- \varepsilon_{0}c_{p, k}^{(i)}E_{k} &= 0.  \\[0pt] 
\mu_0 \partial_t H_k+(\nabla \times \mathbf{E})_k &=0.
\end{empheq}
\end{subequations}
We emphasize our chosen material interpolation in Eq.\,(\ref{Eq:InterpolatedModel}), the parameters $a_{p, k}^{(i)}$ and $c_{p, k}^{(i)}$ do \textit{not} depend on the density itself. In contrast, all fields depend implicitly on $\rho$, and the local functions interpolating the material such as $\varepsilon_{\infty, k}$, $\sigma_{k}$ and $\kappa^{(i)}$ depend directly on the spatial density distribution.\par
We define adjoint fields $\mathbf{\tilde{E}}$, $\mathbf{\tilde{H}}$ and $\tilde{Q}_{p, k}^{(i)}$, for $i =1,2$, $k\in\{1,2,3\}$ and $\forall p \in \{1, \ldots, P_k^{(i)}\}$,
sharing the same properties as the forward fields, i.e. the electric and magnetic adjoint fields are real, and the adjoint auxiliary fields are allowed to be complex. 
We derivate the system of Eqs.\,(\ref{Eq:Maxwells2}) with respect to $\rho$, and multiply Eq.\,(\ref{Eq:Maxwells2}a) by $\tilde{E_k}$, Eq.~(\ref{Eq:Maxwells2}c) by $\tilde{H_k}$, and each of Eqs.\,(\ref{Eq:Maxwells2}b) by a corresponding term $-\kappa^{(i)}\frac{\partial_{t}\tilde{Q}_{p, k}^{(i)}}{\varepsilon_{0} c_{p, k}^{(i)}}$, assuming a non-vanishing parameters $c_{p, k}^{(i)} \neq 0$. Furthermore, we sum over the spatial components and get
\begin{equation}\label{Eq:MaxwellsDerivE}
\begin{split}
&-\sum_{k=1}^{3}\left\{(\nabla \times \mathrm{d}_{\rho}\mathbf{H})_{k}\tilde{E}_k 
+ \varepsilon_{0}(\mathrm{d}_{\rho}\varepsilon_{\infty, k})\tilde{E}_k \partial_{t}E_{k}+ \right. \\[-8pt] 
&\left.  + \varepsilon_{0}\varepsilon_{\infty, k}\tilde{E}_k \partial_{t}(\mathrm{d}_{\rho}E_{k})\right\} \;
\!\!+\!\! \sum_{k=1}^{3}\left\{(\mathrm{d}_{\rho}\sigma_{k}) \tilde{E}_k E_{k}
+ \sigma_{k} \tilde{E}_k (\mathrm{d}_{\rho}E_{k})\right\}+ \\[-8pt] &
+ \sum_{k=1}^{3}\sum_{i \in \{1,2\}}\sum_{p=1}^{P_{k}^{(i)}}2(\mathrm{d}_{\rho}\kappa^{(i)})\tilde{E}_{k}\Re\left\{\partial_{t}Q_{p, k}^{(i)}\right\} +\\[-3pt]  
&+ \sum_{k=1}^{3}\sum_{i \in \{1,2\}}\sum_{p=1}^{P_{k}^{(i)}}2\kappa^{(i)}\tilde{E}_{k}\Re\left\{\partial_{t}(\mathrm{d}_{\rho}Q_{p, k}^{(i)})\right\} = 0,
\end{split}
\end{equation}
\begin{equation}\label{Eq:MaxwellsDerivH}
\begin{split}
\sum_{k=1}^{3}\left\{\mu_0 \tilde{H_k}\partial_t \mathrm{d}_{\rho}H_k+ \tilde{H_k}(\nabla \times \mathrm{d}_{\rho}\mathbf{E})_k\right\} =0.\\
\end{split}
\end{equation}
And for $i =1,2$ and $\forall p \in \{1, \ldots, P_k^{(i)}\}$ we obtain:
\begin{equation}\label{Eq:MaxwellsDerivQ}
\begin{split}
&\sum_{k=1}^{3}\left\{\frac{-\kappa^{(i)}}{\varepsilon_{0} c_{p, k}^{(i)}}\partial_{t}\tilde{Q}_{p, k}^{(i)}\partial_{t}(\mathrm{d}_{\rho}Q_{p, k}^{(i)})+ \right. \\
&\left. +\frac{\kappa^{(i)} a_{p, k}^{(i)}}{\varepsilon_{0} c_{p, k}^{(i)}}\partial_{t}\tilde{Q}_{p, k}^{(i)} (\mathrm{d}_{\rho}Q_{p, k}^{(i)}) +\kappa^{(i)}\partial_{t}\tilde{Q}_{p, k}^{(i)}(\mathrm{d}_{\rho}E_{k})\right\}= 0.
\end{split}
\end{equation}
By addition of the the complex conjugates (denoted as ``c.c.'') of Eqs.\,(\ref{Eq:MaxwellsDerivQ}), and summing over the indices $i$ and $p$, we reduce the equations above to:
\begin{equation}\label{Eq:Qsum}
\begin{split}
\sum_{k=1}^{3}\sum_{i \in \{1,2\}}\sum_{p=1}^{P_{k}^{(i)}}\left\{\frac{-\kappa^{(i)}}{\varepsilon_{0} c_{p, k}^{(i)}}\partial_{t}\tilde{Q}_{p, k}^{(i)}\partial_{t}(\mathrm{d}_{\rho}Q_{p, k}^{(i)}) +\right. \\[-2pt] 
\left.
+\frac{\kappa^{(i)} a_{p, k}^{(i)}}{\varepsilon_{0} c_{p, k}^{(i)}}\partial_{t}\tilde{Q}_{p, k}^{(i)} (\mathrm{d}_{\rho}Q_{p, k}^{(i)})+\mathrm{c.c.}\right\} +\\[-2pt]
+\sum_{k=1}^{3}\sum_{i \in \{1,2\}}\sum_{p=1}^{P_{k}^{(i)}}2\kappa^{(i)}\Re\left\{\partial_{t}\tilde{Q}_{p, k}^{(i)}\right\} (\mathrm{d}_{\rho}E_{k})  &= 0,
\end{split}
\end{equation}
where we used the identities \\ $2\Re\left\{\partial_{t}\tilde{Q}_{p, k}^{(i)}\right\} = \partial_{t}\tilde{Q}_{p, k}^{(i)} + \partial_{t}\tilde{Q}_{p, k}^{*(i)}$, and $\mathrm{d}_{\rho}E^{*}_{k} = \mathrm{d}_{\rho}E_{k}$.\par 
For a better readability, we will waive the symbol ``$\mathrm{d}^3r\,\mathrm{d}t$'' denoting the differential of the variable $(\mathbf{r}, t)$ in all following integral expressions. Integrating over space and time $\Omega \times I$, considering $\rho$ not to be time-dependent, and applying integration by parts in Eqs.\,(\ref{Eq:MaxwellsDerivE}) and (\ref{Eq:MaxwellsDerivH}), while taking the imposed boundary conditions into account, leads to
\begin{equation} \label{Eq:Integr1}
\begin{split}
\int_{\Omega}\!\! \rlap{$\overbrace{\phantom{\left.\sum_{k=1}^{3}\varepsilon_{0}(\mathrm{d}_{\rho}\varepsilon_{\infty, k})\tilde{E}_k E_{k}\right|_0 ^T }}^{=\;0}$}\left. \sum_{k=1}^{3}\varepsilon_{0}(\mathrm{d}_{\rho}\varepsilon_{\infty, k})\tilde{E}_k E_{k}\right|_0 ^T 
\!\!- \!\! \int_{\Omega}\int_{I}\sum_{k=1}^{3}\varepsilon_{0}(\mathrm{d}_{\rho}\varepsilon_{\infty, k})\partial_{t}\tilde{E}_k E_{k}+\\[-5pt]
+\!\!\int_{\Omega}\!\! \rlap{$\overbrace{\phantom{\left.\sum_{k=1}^{3}\varepsilon_{0}\varepsilon_{\infty, k}\tilde{E}_k (\mathrm{d}_{\rho}E_{k})\right|_0 ^T}}^{=\;0}$}\left.\sum_{k=1}^{3}\varepsilon_{0}\varepsilon_{\infty, k}\tilde{E}_k (\mathrm{d}_{\rho}E_{k})\right|_0 ^T 
\!\!-\!\! \int_{\Omega}\int_{I}\sum_{k=1}^{3}\varepsilon_{0}\varepsilon_{\infty, k}\partial_{t}\tilde{E}_k (\mathrm{d}_{\rho}E_{k})+\\
+ \int_{\Omega}\int_{I} \sum_{k=1}^{3}(\mathrm{d}_{\rho}\sigma_{k}) \tilde{E}_k E_{k}
+ \int_{\Omega}\int_{I} \sum_{k=1}^{3}\sigma_{k} \tilde{E}_k (\mathrm{d}_{\rho}E_{k})+\\[-3pt]
+\int_{\Omega}\rlap{$\overbrace{\phantom{\left.\sum_{k=1}^{3}\sum_{i \in \{1,2\}}\sum_{p=1}^{P_{k}^{(i)}}2(\mathrm{d}_{\rho}\kappa^{(i)})\tilde{E}_{k}\Re\left\{Q_{p, k}^{(i)}\right\}\right|_0 ^T}}^{=\;0}$}\left.\sum_{k=1}^{3}\sum_{i \in \{1,2\}}\sum_{p=1}^{P_{k}^{(i)}}2(\mathrm{d}_{\rho}\kappa^{(i)})\tilde{E}_{k}\Re\left\{Q_{p, k}^{(i)}\right\}\right|_0 ^T +\\[-3pt]
- \int_{\Omega}\int_{I} \sum_{k=1}^{3}\sum_{i \in \{1,2\}}\sum_{p=1}^{P_{k}^{(i)}}2(\mathrm{d}_{\rho}\kappa^{(i)})\partial_{t}\tilde{E}_{k}\Re\left\{Q_{p, k}^{(i)}\right\}+\\[-3pt]
+\int_{\Omega}\rlap{$\overbrace{\phantom{\left.\sum_{k=1}^{3}\sum_{i \in \{1,2\}}\sum_{p=1}^{P_{k}^{(i)}}2\kappa^{(i)}\tilde{E}_{k}\Re\left\{(\mathrm{d}_{\rho}Q_{p, k}^{(i)})\right\}\right|_0 ^T}}^{=\;0}$}\left.\sum_{k=1}^{3}\sum_{i \in \{1,2\}}\sum_{p=1}^{P_{k}^{(i)}}2\kappa^{(i)}\tilde{E}_{k}\Re\left\{(\mathrm{d}_{\rho}Q_{p, k}^{(i)})\right\}\right|_0 ^T +\\[-3pt]
- \int_{\Omega}\int_{I}\sum_{k=1}^{3}  \sum_{i \in \{1,2\}}\sum_{p=1}^{P_{k}^{(i)}}2\kappa^{(i)}\partial_{t}\tilde{E}_{k}\Re\left\{(\mathrm{d}_{\rho}Q_{p, k}^{(i)})\right\} +\\[-3pt]
-\int_{\Omega}\rlap{$\overbrace{\phantom{\left.\sum_{k=1}^{3}\partial_{k}(\mathrm{d}_{\rho}\mathbf{H} \times \tilde{\mathbf{E}} )_{k}\right|_0 ^T}}^{=\;0}$}\left.\sum_{k=1}^{3}\partial_{k}(\mathrm{d}_{\rho}\mathbf{H} \times \tilde{\mathbf{E}} )_{k}\right|_0 ^T 
\!\!-\!\! \int_{\Omega}\int_{I}\sum_{k=1}^{3} (\nabla \times \mathbf{\tilde{E}})_{k} (\mathrm{d}_{\rho} H_{k})+\\[-3pt]
+\int_{\Omega}\rlap{$\overbrace{\phantom{\left.\sum_{k=1}^{3}\mu_0 \tilde{H_k} \mathrm{d}_{\rho}H_k\right|_0 ^T}}^{=\;0}$}\left.\sum_{k=1}^{3}\mu_0 \tilde{H_k} \mathrm{d}_{\rho}H_k\right|_0 ^T 
\!\!-\!\! \int_{\Omega}\int_{I} \sum_{k=1}^{3} \mu_0 \partial_{t}\tilde{H_k} (\mathrm{d}_{\rho}H_k) +\\[-3pt]
+\int_{\Omega}\rlap{$\overbrace{\phantom{\left.\sum_{k=1}^{3}\partial_{k}(\mathbf{\tilde{H}} \times \mathrm{d}_{\rho}\mathbf{E})_k\right|_0 ^T}}^{=\;0}$}\left.\sum_{k=1}^{3}\partial_{k}(\mathbf{\tilde{H}} \times \mathrm{d}_{\rho}\mathbf{E})_k\right|_0 ^T 
\!\!\!+\!\! \int_{\Omega}\int_{I} \sum_{k=1}^{3} (\nabla \!\times\! \mathbf{\tilde{H}})_{k}(\mathrm{d}_{\rho}E_k)
=0.
\end{split}
\end{equation}
We do the same  for Eq.\,(\ref{Eq:Qsum}) and obtain:
\begin{equation} \label{Eq:Integr2}
\begin{split}
\int_{\Omega}\rlap{$\overbrace{\phantom{\left.\sum_{k=1}^{3}\sum_{i \in \{1,2\}}\sum_{p=1}^{P_{k}^{(i)}}\left(\frac{-\kappa^{(i)}}{\varepsilon_{0} c_{p, k}^{(i)}}\partial_{t}\tilde{Q}_{p, k}^{(i)}(\mathrm{d}_{\rho}Q_{p, k}^{(i)}) + \mathrm{c.c}\right)\right|_0 ^T}}^{=\;0}$}\left.\sum_{k=1}^{3}\sum_{i \in \{1,2\}}\sum_{p=1}^{P_{k}^{(i)}}\left\{\frac{-\kappa^{(i)}}{\varepsilon_{0} c_{p, k}^{(i)}}\partial_{t}\tilde{Q}_{p, k}^{(i)}(\mathrm{d}_{\rho}Q_{p, k}^{(i)})+ \mathrm{c.c}\right\}\right|_0 ^T +\\
+ \int_{\Omega}\int_{I} \sum_{k=1}^{3} \sum_{i \in \{1,2\}}\sum_{p=1}^{P_{k}^{(i)}}\left\{\frac{\kappa^{(i)}}{\varepsilon_{0} c_{p, k}^{(i)}}\partial^{2}_{t}\tilde{Q}_{p, k}^{(i)}(\mathrm{d}_{\rho}Q_{p, k}^{(i)})+ \right. \\
\left. + \frac{\kappa^{(i)} a_{p, k}^{(i)}}{\varepsilon_{0} c_{p, k}^{(i)}}\partial_{t}\tilde{Q}_{p, k}^{(i)} (\mathrm{d}_{\rho}Q_{p, k}^{(i)}) + \mathrm{c.c.}\right\}+\\
+\int_{\Omega}\int_{I} \sum_{k=1}^{3}\sum_{i \in \{1,2\}}\sum_{p=1}^{P_{k}^{(i)}}2\kappa^{(i)}\Re\left\{\partial_{t}\tilde{Q}_{p, k}^{(i)}\right\} (\mathrm{d}_{\rho}E_{k}) =0.
\end{split}
\end{equation}
By adding Eqs.\,(\ref{Eq:Integr1}) and\,(\ref{Eq:Integr2}), and the integral of the functional derivative in Eq.\,(\ref{Eq:ObjectiveDerivative}) over $\Omega \times I$, we obtain
\begin{equation}\label{Eq:integralRearanged}
\begin{split}
\int_{\Omega}\int_{I}\sum_{k=1}^{3} \Bigg\{\Bigg( (\nabla \times \mathbf{\tilde{H}})_{k} - \varepsilon_{0}\varepsilon_{\infty, k}\partial_{t}\tilde{E}_k  + \sigma_{k} \tilde{E}_k+\qquad   \\
 + 2\sum_{i \in \{1,2\}}\sum_{p=1}^{P_{k}^{(i)}}\kappa^{(i)}\Re\left\{\partial_{t}\tilde{Q}_{p, k}^{(i)}\right\} - \left(\frac{\delta F[\mathbf{E}]}{\delta \mathbf{E}}\right)_{k}\Bigg)           
(\mathrm{d}_{\rho}E_k)\Bigg\}+ \\
 +\int_{\Omega}\int_{I} \sum_{k=1}^{3} \sum_{i \in \{1,2\}}\sum_{p=1}^{P_{k}^{(i)}}\Bigg\{\Bigg(\frac{\kappa^{(i)}}{\varepsilon_{0} c_{p, k}^{(i)}}\partial^{2}_{t}\tilde{Q}_{p, k}^{(i)} + \frac{\kappa^{(i)} a_{p, k}^{(i)}}{\varepsilon_{0} c_{p, k}^{(i)}}\partial_{t}\tilde{Q}_{p, k}^{(i)} + \\ 
 -  \kappa^{(i)}\partial_{t}\tilde{E}_{k}\Bigg)(\mathrm{d}_{\rho}Q_{p, k}^{(i)}) + \mathrm{c.c.}\Bigg\}+\\
 +\int_{\Omega}\int_{I} \sum_{k=1}^{3}\Bigg\{\Bigg( - \mu_0 \partial_{t}\tilde{H_k} - (\nabla \times \mathbf{\tilde{E}})_{k} \Bigg)(\mathrm{d}_{\rho}H_k)\Bigg\}+\\[2pt]
 +\int_{\Omega}\int_{I}\frac{\delta F[\mathbf{E}]}{\delta \rho} - \int_{\Omega}\int_{I}\sum_{k=1}^{3}\Bigg\{ \varepsilon_{0}(\mathrm{d}_{\rho}\varepsilon_{\infty, k})\partial_{t}\tilde{E}_k E_{k} +  \\
-(\mathrm{d}_{\rho}\sigma_{k}) \tilde{E}_k E_{k} + 2 \sum_{i \in \{1,2\}}\sum_{p=1}^{P_{k}^{(i)}}(\mathrm{d}_{\rho}\kappa^{(i)})\partial_{t}\tilde{E}_{k}\Re\left\{Q_{p, k}^{(i)}\right\}\Bigg\}
 =0.
\end{split}
\end{equation}
We note that the Eq.\,(\ref{Eq:integralRearanged}) is satisfied, if the following equations hold $\forall (\mathbf{r},t) \in \Omega \times I$ and each spatial component $k \in \{1,2,3\}$:
\begin{subequations}\label{Eq:AlmostAdjoint}
  \begin{empheq}[]{align}
\!\!(\nabla \times \mathbf{\tilde{H}})_{k} - \varepsilon_{0}\varepsilon_{\infty, k}\partial_{t}\tilde{E}_k  + \sigma_{k} \tilde{E}_k+ \nonumber \\
 + 2 \!\!\!\sum_{i \in \{1,2\}} \!\sum_{p=1}^{P_{k}^{(i)}}\kappa^{(i)}\Re\left\{\partial_{t}\tilde{Q}_{p, k}^{(i)}\right\}&= \left(\frac{\delta F[\mathbf{E}]}{\delta \mathbf{E}}\right)_{k},\\
\!\!\text{For } i =1,2 \text{ and }  
 \forall p \in \{1, \ldots, P_k^{(i)}\}\!:  & \nonumber \\ \partial_{t}\tilde{Q}_{p, k}^{(i)}+a_{p, k}^{(i)}\tilde{Q}_{p, k}^{(i)}- \varepsilon_{0}c_{p, k}^{(i)}\tilde{E}_{k} &= 0,\\[8pt]
\mu_0 \partial_{t}\tilde{H_k} +(\nabla \times \mathbf{\tilde{E}})_{k} &=0,
\end{empheq}
\end{subequations}
and if $\forall \mathbf{r}\in \Omega$ the gradient of the objective $\nabla_{\rho}F[\mathbf{E}]$ defined  as
\begin{equation}\label{Eq:IntegralKernel}
\begin{split}
\nabla_{\rho}F[\mathbf{E}]:=\int_{I}\frac{\delta F[\mathbf{E}]}{\delta \rho},
\end{split}
\end{equation}
satisfies the equation
\begin{equation}\label{Eq:Gradients2}
\begin{split}
\nabla_{\rho}F[\mathbf{E}]=&\phantom{+}\int_{I}\sum_{k=1}^{3}\varepsilon_{0}(\mathrm{d}_{\rho}\varepsilon_{\infty, k})\partial_{t}\tilde{E}_k E_{k} +\\
&-\int_{I}\sum_{k=1}^{3}(\mathrm{d}_{\rho}\sigma_{k}) \tilde{E}_k E_{k}+ \\
 &+\int_{I}\sum_{k=1}^{3}\sum_{i \in \{1,2\}}\sum_{p=1}^{P_{k}^{(i)}}2(\mathrm{d}_{\rho}\kappa^{(i)})\partial_{t}\tilde{E}_{k}\Re\left\{Q_{p, k}^{(i)}\right\}.
\end{split}
\end{equation}
Now, we perform transformations of the fields in Eqs.\,(\ref{Eq:AlmostAdjoint}) to obtain a system of Maxwell equations for the adjoint system. First, we reverse the time and change the sign of the magnetic field $\mathbf{\tilde{H}}$  and the currents $\tilde{Q}_{p, k}^{(i)}$ accordingly, i.e.
\begin{subequations}\label{Eq:Tranformations}
  \begin{empheq}[]{align}
&\mathbf{\tilde{H}}(t)\rightarrow-\mathbf{\tilde{H}}(\tau),\\
&\tilde{Q}_{p, k}^{(i)}(t)\rightarrow-\tilde{Q}_{p, k}^{(i)}(\tau),  \\ 
&  \text{for } k\in\{1,2,3\},\;i =1,2 \text{ and } \forall p \in \{1, \ldots, P_k^{(i)}\} \nonumber.
\end{empheq}
\end{subequations}
where $\tau:= T - t$ denotes the time-reversed variable. Furthermore, we require vanishing fields for $\tau=0$. If we now apply the chain rule for the time derivatives of all time reversed functions, we finally obtain the adjoint system which holds $\forall (\mathbf{r}, \tau) \in \Omega \times [0, T]$,
\begin{subequations}\label{Eq:AdjointSystem}
  \begin{empheq}[]{align}
\!-(\nabla \times \mathbf{\tilde{H}})_{k} + \varepsilon_{0}\varepsilon_{\infty, k}\partial_{\tau}\tilde{E}_k  \!+ \sigma_{k} \tilde{E}_k + \nonumber\\ 
+ 2\! \!\!\!\sum_{i \in \{1,2\}} \!\sum_{p=1}^{P_{k}^{(i)}}\kappa^{(i)}\Re\left\{\partial_{\tau}\tilde{Q}_{p, k}^{(i)}\right\}&= \!\left(\overleftarrow{\frac{\delta F[\mathbf{E}]}{\delta \mathbf{E}}}\right)_{\!\!k}\!\!,\\
\text{For } i =1,2 \text{ and } \forall p \in \{1, \ldots, P_k^{(i)}\}: \nonumber \\
\partial_{\tau}\tilde{Q}_{p, k}^{(i)}-a_{p, k}^{(i)}\tilde{Q}_{p, k}^{(i)}- \varepsilon_{0}c_{p, k}^{(i)}\tilde{E}_{k} &= 0,\\
\mu_0 \partial_{\tau}\tilde{H_k} +(\nabla \times \mathbf{\tilde{E}})_{k} &=0.
\end{empheq}
\end{subequations}
Here, the symbol ``$\overleftarrow{}$'' over the adjoint source term denotes the time-reversal transformation. 
Applying these transformation on the gradients in Eq.\,(\ref{Eq:Gradients2}), leads to
\begin{equation}\label{Eq:Gradients1}
\begin{split}
\nabla_{\rho}F[\mathbf{E}]=&-\int_{I}\sum_{k=1}^{3}\varepsilon_{0}(\mathrm{d}_{\rho}\varepsilon_{\infty, k})\partial_{\tau}\tilde{E}_k \overleftarrow{E}_{k} +\\
&-\int_{I}\sum_{k=1}^{3}(\mathrm{d}_{\rho}\sigma_{k}) \tilde{E}_k \overleftarrow{E}_{k} +\\
 &-\int_{I}\sum_{k=1}^{3}\sum_{i \in \{1,2\}}\sum_{p=1}^{P_{k}^{(i)}}2(\mathrm{d}_{\rho}\kappa^{(i)})\partial_{\tau}\tilde{E}_{k}\Re\left\{\overleftarrow{Q}_{p, k}^{(i)}\right\}.
\end{split}
\end{equation}
This equation is equivalent to
\begin{equation}\label{Eq:Gradients3}
\begin{split}
\nabla_{\rho}F[\mathbf{E}]=&-\int_{I}\sum_{k=1}^{3}\varepsilon_{0}(\mathrm{d}_{\rho}\varepsilon_{\infty, k})\partial_{\tau}\tilde{E}_k \overleftarrow{E}_{k} +\\
&-\int_{I}\sum_{k=1}^{3}(\mathrm{d}_{\rho}\sigma_{k}) \tilde{E}_k \overleftarrow{E}_{k} +\\
 &+\int_{I}\sum_{k=1}^{3}\sum_{i \in \{1,2\}}\sum_{p=1}^{P_{k}^{(i)}}2(\mathrm{d}_{\rho}\kappa^{(i)})\tilde{E}_{k}\Re\left\{\partial_{\tau}\overleftarrow{Q}_{p, k}^{(i)}\right\},
\end{split}
\end{equation}
if we again apply integration by parts on the last term and taking the imposed boundary conditions into account.

\section{Instantaneous electric power dissipation based on the CCPR model} 
Here, we derive the expression for the density-dependent instantaneous electric power dissipation based on a linear mixing scheme of background and design material (Eq.~(\ref{Eq:CCPR_model_mixed})), whose time average will serve as our objective function. We follow a similar procedure as in Ref.~\cite{Shin:12}, which is therein restricted to materials described by the Lorentz model. \newline 
The general Poynting theorem in a source-free medium reads
\begin{equation}\label{Eq:PoyntingTheorem_1}
-\oint_{\partial\Omega_{\mathrm{o}}}(\mathbf{E} \times \mathbf{H}) \cdot \mathrm{d} \mathbf{a}=\int_{\Omega_{\mathrm{o}}}\left(\mathbf{E} \cdot \frac{\partial \mathbf{D}}{\partial t}+\mathbf{H} \cdot \frac{\partial \mathbf{B}}{\partial t}\right) \mathrm{d}^3 r,
\end{equation}
and using the ansatz,
\begin{equation}\label{Eq:PoyntingTheorem_2}
-\oint_{\partial\Omega_{\mathrm{o}}}\!\! (\mathbf{E} \times \mathbf{H}) \cdot \mathrm{d} \mathbf{a}=\!\!\int_{\Omega_{\mathrm{o}}} \!\!\!\left(\frac{\partial u_e}{\partial t}+\frac{\partial u_m}{\partial t}\right) \mathrm{d}^3 r+\int_{\Omega_{\mathrm{o}}}\!\! \left(q_e\!+\!q_m\right)  \mathrm{d}^3 r,
\end{equation}
where $q_e(t)$ and $q_m(t)$ are the instantaneous electric and magnetic power dissipation densities, and $u_e(t)$ and $u_m(t)$ are assumed to be the instantaneous electric and magnetic energy densities, respectively.\newline
In the following, just a single pole, and an isotropic, diagonal permittivity tensor for both materials is considered for reasons of better readability, but the derivations can easily be extended to the anisotropic case and multiple poles. According to the linear mixed permittivities based on the CCPR model described by a density $\rho$, we obtain the material-interpolated equation
\begin{equation}
\partial_t \mathbf{D}=\nabla \times \mathbf{H}=\varepsilon_0 \varepsilon_{\infty} \partial_t \mathbf{E}+\sigma \mathbf{E}+2 \sum_{i \in \{1,2\}} \kappa^{(i)} \Re\left\{\partial_t \mathbf{Q}^{(i)}\right\},
\end{equation}
such that
\begin{equation}\label{Eq:Dens+Dissip_1stForm}
\mathbf{E}\cdot\partial_t \mathbf{D}=
\dfrac{\varepsilon_0 \varepsilon_{\infty}}{2} \partial_t (\mathbf{E}^{2})+\sigma \mathbf{E}^{2}+2 \sum_{i \in \{1,2\}}  \kappa^{(i)} \mathbf{E}\cdot\Re\left\{\partial_t \mathbf{Q}^{(i)}\right\}.
\end{equation}
We recall that the prefactors $\varepsilon_{\infty} := \varepsilon_{\infty}(\rho)$,  $\sigma := \sigma(\rho)$, and $\kappa^{(i)}:=\kappa^{(i)}(\rho)$ are density-dependent.\newline 
Using the relation between electric and auxiliary fields from Eq.~(\ref{Eq:Maxwells2}.b), assuming $c^{(i)} \neq 0$, for $i=1,2$,
\begin{equation}
\mathbf{E} = \frac{\partial_t \mathbf{Q}^{(i)}}{\varepsilon_{0}c^{(i)}} - \frac{a^{(i)}\mathbf{Q}^{(i)}}{\varepsilon_{0}c^{(i)}},
\end{equation}
where $\mathbf{E}$ is supposed to be real, it follows:
\begin{equation}
\begin{split}
\mathbf{E}\cdot\Re\left\{\partial_t \mathbf{Q}^{(i)}\right\} & = \Re\left\{\mathbf{E}\cdot\partial_t \mathbf{Q}^{(i)}\right\}  \\ & =\Re\left\{\frac{(\partial_{t}\mathbf{Q}^{(i)})^2}{\varepsilon_{0}c^{(i)}} \!-\!\frac{a^{(i)}\partial_t (\mathbf{Q}^{(i)})^2}{2\varepsilon_{0}c^{(i)}}\right\},
\end{split}
\end{equation}
and Eq.~(\ref{Eq:Dens+Dissip_1stForm}) becomes
\begin{equation}\label{Eq:Dens+Dissip_2ndForm}
\begin{split}
\mathbf{E}\cdot\partial_t \mathbf{D}=&\,
\dfrac{1}{2}\varepsilon_0 \varepsilon_{\infty} \partial_t (\mathbf{E}^{2})+\sigma \mathbf{E}^{2} + \\& +2 \sum_{i \in \{1,2\}} \kappa^{(i)} \Re\left\{\frac{(\partial_{t}\mathbf{Q}^{(i)})^2}{\varepsilon_{0}c^{(i)}}-\frac{a^{(i)}\partial_t (\mathbf{Q}^{(i)})^2}{2\varepsilon_{0}c^{(i)}}\right\} +\\
=&\,\dfrac{1}{2}\varepsilon_0 \varepsilon_{\infty} {\color{blue}\partial_{t}} (\mathbf{E}^{2})+\sigma \mathbf{E}^{2} +2\!\!\! \sum_{i \in \{1,2\}}  \kappa^{(i)} \Re\left\{\frac{(\partial_{t}\mathbf{Q}^{(i)})^2}{\varepsilon_{0}c^{(i)}}\right\} + \\ 
& -{\color{blue}\partial_{t}} \sum_{i \in \{1,2\}} \kappa^{(i)}\Re\left\{\frac{a^{(i)}(\mathbf{Q}^{(i)})^2}{\varepsilon_{0}c^{(i)}}\right\} +\\
=& \,\sigma \mathbf{E}^{2}+2 \sum_{i \in \{1,2\}}  \kappa^{(i)} \Re\left\{\frac{(\partial_{t}\mathbf{Q}^{(i)})^2}{\varepsilon_{0}c^{(i)}}\right\} + \\ 
&+{\color{blue}\partial_{t}}\left(\dfrac{1}{2}\varepsilon_0 \varepsilon_{\infty} \mathbf{E}^{2} -\sum_{i \in \{1,2\}} \kappa^{(i)}\Re\left\{\frac{a^{(i)}(\mathbf{Q}^{(i)})^2}{\varepsilon_{0}c^{(i)}}\right\}\right)
\end{split}
\end{equation}
Identifying $\mathbf{E}\cdot\partial_t \mathbf{D}={\color{blue}\partial_{t}} u_e + q_e$ from Eqs.~(\ref{Eq:PoyntingTheorem_1}) and (\ref{Eq:PoyntingTheorem_2}), the instantaneous electric power dissipation density must be
\begin{equation}\label{Eq:InstantanousDissipation}
q_e([\rho];t) = \sigma(\rho) \mathbf{E}^{2} + 2\sum_{i \in \{1,2\}}  \kappa^{(i)}(\rho) \Re\left\{\frac{(\partial_{t}\mathbf{Q}^{(i)})^2}{\varepsilon_{0}c^{(i)}}\right\}.
\end{equation}

\subsection{The dc component of the electric instantaneous power dissipation}
In the time-harmonic case, the instantaneous electric power dissipation density in Eq.~(\ref{Eq:InstantanousDissipation})  can be separated in a dc and oscillating ac component \cite{Shin:12}. To measure the broadband performance of an optimized device in the steady state regarding the time-average of that dissipation density, we must extract its dc component. \newline
Consider a time harmonic field $\mathbf{E}(t):= \Re\{\mathbf{\hat{E}}_0 e^{j\omega_{0} t}\}$ oscillating with a frequency $\omega_{0}$, where $\hat{\mathbf{E}}_{0}$ itself is frequency independent. The Fourier transform yields
\begin{equation}
\hat{\mathbf{E}}(\omega)=\frac{1}{2}\left(\hat{\mathbf{E}}_{0}\delta\left(\omega_0-\omega\right)+\hat{\mathbf{E}}_{0}^{*} \delta\left(\omega_0+\omega\right)\right).
\end{equation}
The auxillary field components in frequency domain based on the CCPR model in Eq.~(\ref{Eq:CCPR_model}) are related to the electric field components via \cite{material}
\begin{equation}
\hat{\mathbf{Q}}(\omega)=\frac{\varepsilon_0 c}{j \omega-a} \hat{\mathbf{E}}(\omega).
\end{equation}
Thus, based on the time-harmonic field, the auxillary field in time domain and its derivative read
\begin{equation}
\begin{split}
\mathbf{Q}(t)&=\int_{-\infty}^{+\infty} \tilde{\mathbf{Q}}(\omega) e^{j \omega t} d \omega \\
&=\frac{1}{2}\left(\frac{\varepsilon_0 c}{j \omega_0-a} \hat{\mathbf{E}}_{0} e^{j \omega_0 t}+\frac{\varepsilon_0 c}{-j \omega_0-a} \hat{\mathbf{E}}_{0}^{*} e^{-j \omega_0 t}\right),
\end{split}
\end{equation}
\begin{equation}
\partial_{t} \mathbf{Q}(t)=\frac{1}{2}\left(\frac{j \omega_0 \varepsilon_0 c}{j \omega_0-a} \hat{\mathbf{E}}_{0} e^{j \omega_0 t}+\frac{-j \omega_0 \varepsilon_0 c}{-j \omega_0-a} \hat{\mathbf{E}}_{0}^{*} e^{-j \omega_0 t}\right).
\end{equation}
Substituting $\partial_{t} \mathbf{Q}^{(i)}$ into  Eq.~(\ref{Eq:InstantanousDissipation}), for both background and design material $i=1,2$, yields an expression for both the ac and dc component (see Ref~\cite{Shin:12}) of the \textit{density-dependent} electric power dissipation. Considering the time-average of the dissipation as the objective we later aim to optimize, we are only interested in the dc component. Using the identity
\begin{equation}
\Re\{\mathbf{W}\}^2=\frac{1}{4}\left(\mathbf{W}+\mathbf{W}^*\right)^2=\frac{1}{2}|\mathbf{W}|^2+\frac{1}{2} \Re\left\{\mathbf{W}^2\right\},
\end{equation}
that holds for any complex vector field $\mathbf{W}$, and neglecting all the terms consisting $e^{\pm 2j\omega t}$, we obtain the expression for the dc component of the electric power dissipation as
\begin{equation}\label{Eq:Dissipation_Freq}
\begin{split}
&q_{e}^{\mathrm{dc}}([\rho];\omega_{0}) = \\ 
&\!\!\left(\!\frac{1}{2}\sigma(\rho)  \!+\!\!\!\!\! \sum_{i \in \{1,2\}}^2\!\!\!\!\! \kappa^{(i)}(\rho)\Re\left\{\frac{\omega_{0}^{2} \varepsilon_{0} c^{(i)}}{(j\omega_{0} - a^{(i)})(-j\omega_{0} - a^{(i)})}\right\}\!\!\right)\! \|\hat{\mathbf{E}}_{0}\|^{2}\!.
\end{split}
\end{equation}
We can express the result in terms of the imaginary part of the (density-dependent) permittivity $\varepsilon$ in \mbox{Eqs.~(\ref{Eq:CCPR_model})-(\ref{Eq:InterpolatedModel})},
\begin{equation}\label{Eq:Landau_equivalenz}
\begin{split}
& q_{e}^{\mathrm{dc}}([\rho];\omega_{0})= \\ 
=&\frac{\sigma}{2} + \sum_{i \in \{1,2\}}\Re\left\{\frac{\omega_{0}^{2} \varepsilon_{0} c^{(i)}}{(j\omega_{0} - a^{(i)})(-j\omega_{0} - a^{(i)})}\right\}\|\hat{\mathbf{E}}_{0}\|^{2} \\
=& \frac{\sigma}{2}  -\varepsilon_{0}\omega_{0} \!\!\!\sum_{i \in \{1,2\}} \!\!\!\Re\left\{\frac{1}{2j}\left( \frac{c^{(i)}}{j\omega_{0} - a^{(i)}}  -\! \frac{c^{(i)}}{-j\omega_{0} - a^{(i)}} \right)\!\right\} \!\|\hat{\mathbf{E}}_{0}\|^{2}\\
=& \frac{\sigma}{2}  -\frac{\varepsilon_{0}\omega_{0}}{2} \!\!\!\sum_{i \in \{1,2\}} \frac{1}{2j}\Bigg\{\left( \frac{c^{(i)}}{j\omega_{0} - a^{(i)}}  + \frac{c^{(i)*}}{\phantom{-}j\omega_{0} - a^{(i)*}}\right) + \\ 
&  -  \mathrm{c.c.}\Bigg\} \|\hat{\mathbf{E}}_{0}\|^{2}\\
\!\! = & -\frac{1}{2}\varepsilon_{0}\,\omega_{0}\;\frac{1}{2j} \!\Bigg\{\frac{\sigma}{j\varepsilon_{0}\omega_0} \!+\!\!\!\!\! \sum_{i \in \{1,2\}}\!\!\! \left( \frac{c^{(i)}}{j\omega_{0} - a^{(i)}}  +\! \frac{c^{(i)*}}{\phantom{-}j\omega_{0} - a^{(i)*}}\!\right) \!\!+\\  & - \mathrm{c.c.}\Bigg\}\|\hat{\mathbf{E}}_{0}\|^{2} \\
=& -\frac{1}{2}\varepsilon_{0}\,\omega_{0} \,\Im\left\{ \varepsilon([\rho];\omega_{0}) \right\}\|\hat{\mathbf{E}}_{0}\|^{2}.
\end{split}
\end{equation}
From that we observe, that for a homogenous medium, Eq.~(\ref{Eq:Dissipation_Freq}) reduces to the model-independent expression derived by Landau und Lifschitz in Ref.~\cite{landau1995electrodynamics}, considering a $e^{j\omega t}$ time-dependence and the convention \mbox{$\varepsilon= \varepsilon^{\prime} - j\varepsilon^{\prime\prime}$}.\newline \newline
For the anisotropic case $\varepsilon:=\operatorname{diag}\left(\varepsilon_1, \varepsilon_2, \varepsilon_3\right)$ with multiple poles per component, we must consider an auxiliary field for each pole of the corresponding component of the material, i.e., 
\begin{equation}
Q_{p, k}^{(i)}:= \frac{1}{2}\left(\frac{\varepsilon_{0}c^{(i)}_{p,k}}{j \omega - a^{(i)}_{p,k}}\hat{E}_{k}e^{j\omega t} + \frac{ \varepsilon_{0}c^{(i)}_{p,k}}{-j \omega - a^{(i)}_{p,k}}\hat{E}^{*}_{k}e^{-j\omega t} \right).  
\end{equation}
Thus, the dissipation formula in Eq.~(\ref{Eq:Dissipation_Freq}) must be modified to
\begin{equation}\label{Eq:Dissipation_Freq_with_poles}
\begin{split}
&q_{e}^{\mathrm{dc}}([\rho];\omega_{0})=\sum_{k \in \{1,2,3\}}\Bigg(\frac{1}{2}\sigma_k(\rho) \,+  \\ 
 &  + \!\!\!\! \sum_{i \in \{1,2\}} \!\sum_{p=1}^{P_{k}^{(i)}}\! \kappa^{(i)}(\rho)\Re\left\{\frac{\omega_{0}^{2} \varepsilon_{0} c^{(i)}_{p,k}}{(j\omega_{0} - a^{(i)}_{p,k})(-j\omega_{0} - a^{(i)}_{p,k})}\right\}\!\Bigg)\! |\hat{E}_{0,k}|^{2}.
\end{split}
\end{equation}

\section{Sensitivity analysis for maximizing the electric power dissipation}\label{sec:adjoint_powerdissipation}
Based on the instantaneous electric power dissipation derived in the previous section, we are going to deduce an adjoint scheme for the time-domain topology optimization for maximizing (or minimizing) the electric dissipated power over a time $T$ in an observation volume $\Omega_{o}$. The objective function reads
\begin{equation}\label{Eq:ObjectiveEnerg&Dissipation}
F:= B\,\int_{\Omega_o} \int_0^T q_e([\rho];t) \,\mathrm{d}t \mathrm{d}r^3. 
\end{equation}
 $T$ denotes the time period until the fields are decayed after the excitation of a light pulse, which is injected outside the design domain $\Omega$. We consider the observation region to be located in the design region, i.e., $\Omega_{o} \subseteq \Omega$. $B$ is an arbitrary constant (in units of $T^{-1}$), and $q_e([\rho];t)$ is the instantaneous electric power dissipation from Eq.~(\ref{Eq:InstantanousDissipation}). Here, too, we initially assume isotropic media for background and design material, which are described by a single pole for reasons of readability in the following. The extension to the anisotropic case with multiple poles will be given later.\newline 
We recall that in Eq.~(\ref{Eq:ObjectiveEnerg&Dissipation}) all quantities depend on the density, except $\varepsilon_{0}$, $a^{(i)}$ and $c^{(i)}$, such that the functional derivative of $F$ with respect to the $\rho$ reads
\begin{figure}[t!]
    \centering
    \includegraphics[width=0.4\linewidth]{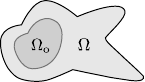}
    \caption{\raggedright Schematic illustration of sets $\Omega$ and $\Omega_o$. The objective function $F$ is defined on a subset $\Omega_o$ lying in the design region $\Omega$. The density can vary in both regions during the optimization iterations. Background and design material and their interpolation is considered to be the same in both regions.}
    \label{Fig:Sets}
\end{figure}
\begin{equation}\label{Eq:DerivativeEnergy+Dissipation}
\begin{split}
\frac{\delta F}{\delta \rho}&=B\Bigg(\textcolor{Green}{2\sigma\mathbf{E}(\mathrm{d}_{\rho}\mathbf{E})} \textcolor{blue}{\,+\, (\mathrm{d}_{\rho}\sigma) \mathbf{E}^{2}}+  \\ 
 &\textcolor{purple}{+ 4\sum_{i \in \{1,2\}}  \kappa^{(i)} \Re\left\{\frac{\partial_{t}\mathbf{Q}^{(i)}}{\varepsilon_{0}c^{(i)}}\partial_t(\mathrm{d}_{\rho}\mathbf{\mathbf{Q}^{(i)}})\right\}+} \\ 
 & \textcolor{blue}{+ 2 \sum_{i \in \{1,2\}} (\mathrm{d}_{\rho}\kappa^{(i)}) \Re\left\{\frac{(\partial_{t}\mathbf{Q}^{(i)})^2}{\varepsilon_{0}c^{(i)}}\right\}}\Bigg)
\end{split}
\end{equation}
Following the derivations of the first section, it can be seen from Eq.~(\ref{Eq:integralRearanged}) that the time reversal of the terms marked in green in the above equation form the adjoint source in Eq.~(\ref{Eq:AdjointSystem}.a), and the time reversal of the terms marked in blue are additional gradient terms in Eq.~(\ref{Eq:Gradients1})/(\ref{Eq:Gradients2}), with corresponding weighting factor $B$.\newline
The terms marked in purple form the adjoint source for the adjoint auxiliary equations in Eq~(\ref{Eq:AdjointSystem}.b), which can be seen as follows: \newline 
We denote the Eq.~(\ref{Eq:AlmostAdjoint}.b) for $i=1,2$ as
\begin{equation}
\mathcal{\tilde{A}}^{(i)}:=\partial_{t}\mathbf{\tilde{Q}}^{(i)}+a^{(i)}\mathbf{\tilde{Q}}^{(i)}- \varepsilon_{0}c^{(i)}\mathbf{\tilde{E}}.
\end{equation}
 Then, the second term in Eq.~(\ref{Eq:integralRearanged}) can be written as
 \begin{equation}\label{Eq:term2New}
\int_{\Omega}\int_{I}  \sum_{i \in \{1,2\}}\left\{\frac{\kappa^{(i)}\partial_t\mathcal{\tilde{A}}^{(i)}}{\varepsilon_{0} c^{(i)}}(\mathrm{d}_{\rho}\mathbf{Q}^{(i)}) + \mathrm{c.c.}\right\}.
 \end{equation}
 Integrating 2nd purple term in Eq.~(\ref{Eq:DerivativeEnergy+Dissipation}) over space and time, using integration by parts and reminding that\newline $\mathrm{d}_{\rho}\mathbf{Q}^{(i)}(0)=\mathrm{d}_{\rho}\mathbf{Q}^{(i)}(T)=0$, leads to 
 \begin{equation}
  \begin{split}
 \int_{\Omega}\int_{I} \sum_{i \in \{1,2\}} \kappa^{(i)} \left\{{\color{purple}\frac{\partial_{t}\mathbf{Q}^{(i)}}{\varepsilon_{0}c^{(i)}}\partial_t(\mathrm{d}_{\rho}\mathbf{Q}^{(i)}) + c.c.}\right\}= \\ - \int_{\Omega}\int_{I} \sum_{i \in \{1,2\}}\kappa^{(i)} \left\{{\color{purple}\frac{\partial^2_{t}\mathbf{Q}^{(i)}}{\varepsilon_{0}c^{(i)}}(\mathrm{d}_{\rho}\mathbf{Q}^{(i)}) + c.c.}\right\},
 \end{split}
 \end{equation}
 and thus, Eq.~(\ref{Eq:term2New}) can be extended to
 \begin{equation}
 \begin{split}
\int_{\Omega_{o}}\int_{I} \! \sum_{i \in \{1,2\}}\!\!\left\{\frac{\kappa^{(i)}}{\varepsilon_{0}c^{(i)}}\left(\partial_t\mathcal{\tilde{A}}^{(i)}{\color{purple}+B\,2\partial^2_{t}\mathbf{Q}^{(i)}}\right)(\mathrm{d}_{\rho}\mathbf{Q}^{(i)})  + \mathrm{c.c.}\right\},
\end{split}
 \end{equation}
taking the weighting factor $B$ from Eq.~(\ref{Eq:DerivativeEnergy+Dissipation}) into account. We emphasize that the integral above is explicitly defined on $\Omega_{o}$, since the objective itself is defined on $\Omega_{o}\subseteq\Omega$.\newline
Enforcing again that $\mathcal{\tilde{A}}^{(i)}+B2\partial_{t}\mathbf{Q}^{(i)}=0$, the Eq.~(\ref{Eq:AlmostAdjoint}.b) for spatial values lying in $\Omega_{o}$ gets modified to
\begin{equation}
\partial_{t}\mathbf{\tilde{Q}}^{(i)}+a^{(i)}\mathbf{\tilde{Q}}^{(i)}- \varepsilon_{0}c^{(i)}\mathbf{\tilde{E}}={\color{purple}-B\,2\partial_{t}\mathbf{Q}^{(i)}}.   
\end{equation}
Performing the time reversal transformation for the fields as in Eqs.~(\ref{Eq:Tranformations}) 
leads to an adjoint system of auxiliary equations as in Eq.~(\ref{Eq:AdjointSystem}), but with an additional source term
\begin{equation}
\partial_{\tau}\mathbf{\tilde{Q}}^{(i)}-a^{(i)}\mathbf{\tilde{Q}}^{(i)}- \varepsilon_{0}c^{(i)}\mathbf{\tilde{E}}={\color{purple}+B\,2\partial_{\tau}\overleftarrow{\mathbf{Q}}^{(i)}}.  
\end{equation}
where the additional source term $B\,2\partial_{\tau}\overleftarrow{\mathbf{Q}}^{(i)}$ comes from the electric dissipated power $q_e$.\newline
We denote the adjoint source terms comming power dissipation and being injected into the observation region $\Omega_{o}$ as
\begin{equation}\label{Eq:Adjoint_Sources}
\begin{split}
\textcolor{Green}{\mathbf{S}_{E}:=}& \textcolor{Green}{\phantom{-}2B\sigma\overleftarrow{\mathbf{E}}},\\
\textcolor{purple}{\mathbf{S}_{\partial_{\tau}Q}^{(i)}:=}&\textcolor{purple}{\phantom{-}2 B \partial_{\tau}\overleftarrow{\mathbf{Q}}^{(i)}}, \;\, \text{for}\; i=1,2.\; 
\end{split}
\end{equation}
Then, the adjoint system of Maxwell's equations in $\Omega_{o}$ reads
\begin{subequations}\label{Eqs:Dissipation}
  \begin{empheq}[]{align}
-\nabla \times \mathbf{\tilde{H}}+\varepsilon_0 \varepsilon_{\infty} \partial_t \mathbf{\tilde{E}} +\sigma \mathbf{\tilde{E}} +  \qquad\qquad\nonumber \\ + 2 \!\!\sum_{i \in \{1,2\}} \!\!\kappa^{(i)} \Re\left\{\partial_t \mathbf{\tilde{Q}}^{(i)}\!\right\}&\!=\!\textcolor{Green}{\mathbf{S}_{E}},\\[0pt]
\text{For } i =1,2:\; \partial_{\tau}\mathbf{\tilde{Q}}^{(i)}\!-\!a^{(i)}\mathbf{\tilde{Q}}^{(i)}\!- \varepsilon_{0}c^{(i)}\mathbf{\tilde{E}}& \!=\!\textcolor{purple}{\mathbf{S}_{\partial_{\tau}Q}^{(i)}},\\[8pt]
\mu_0 \partial_{\tau}\mathbf{\tilde{H}} +(\nabla \times \mathbf{\tilde{E}}) &=0,
\end{empheq}
\end{subequations}
and the gradients of the objective with respect to the density as in is Eq.~(\ref{Eq:Gradients2}) must be extended and can be expressed as
\begin{equation}
\nabla_{\rho}F:= \mathcal{A}_{|\rho\in\Omega}+ \textcolor{blue}{\mathcal{B}_{|\rho\in\Omega_{o}}},
\end{equation}
with
\begin{subequations}\label{Eq:Gradients2_Diss}
 \begin{empheq}[]{align}
\mathcal{A}_{|\rho\in\Omega}:=&\quad\;\;\,\int_{I}\varepsilon_{0}(\mathrm{d}_{\rho}\varepsilon_{\infty})\partial_{t}\mathbf{\tilde{E}} \cdot \mathbf{E} +\nonumber \\ 
&-\int_{I}(\mathrm{d}_{\rho}\sigma) \mathbf{\tilde{E}} \cdot \mathbf{E} + \nonumber \\ 
& +\int_{I}\sum_{i \in \{1,2\}}2(\mathrm{d}_{\rho}\kappa^{(i)})\partial_{t}\mathbf{\tilde{E}}\cdot\Re\left\{\mathbf{Q}^{(i)}\right\}, \\
 \textcolor{blue}{\mathcal{B}_{|\rho\in\Omega_{o}}:=}&\textcolor{blue}{\;\;\;B\int_{I} (\mathrm{d}_{\rho}\sigma) \mathbf{E}^{2}} +\nonumber \\  
& \textcolor{blue}{\;+B\int_{I}\sum_{i \in \{1,2\}}2(\mathrm{d}_{\rho}\kappa^{(i)}) \Re\left\{\frac{(\partial_{t}\mathbf{Q}^{(i)})^2}{\varepsilon_{0}c^{(i)}}\right\}},
\end{empheq}
\end{subequations}
following the same steps leading to Eq.~(\ref{Eq:integralRearanged}), taking the additional terms from Eq.~(\ref{Eq:DerivativeEnergy+Dissipation}) (marked in blue) defined on $\Omega_{o}$ into account. The subindices for $\mathcal{A}$ and $\mathcal{B}$ denote, that the gradient terms are only defined on the corresponding subsets (see Fig.~\ref{Fig:Sets}). Applying the time reversal transformations as in Eq.~(\ref{Eq:Tranformations}), the gradient terms transform to 
\begin{subequations}\label{Eq:GradientsTimeReversal_Diss}
\begin{empheq}[]{align}
\mathcal{A}_{|\rho\in\Omega}:=&\;\;-\int_{I}\varepsilon_{0}(\mathrm{d}_{\rho}\varepsilon_{\infty})\partial_{\tau}\mathbf{\tilde{E}} \cdot \overleftarrow{\mathbf{E}} +\nonumber \\
 & -\int_{I}(\mathrm{d}_{\rho}\sigma) \mathbf{\tilde{E}} \cdot \overleftarrow{\mathbf{E}} +\nonumber \\ 
 & -\int_{I}\sum_{i \in \{1,2\}}2(\mathrm{d}_{\rho}\kappa^{(i)})\partial_{\tau}\mathbf{\tilde{E}}\cdot\Re\left\{\overleftarrow{\mathbf{Q}}^{(i)}\right\}, \\
 \textcolor{blue}{\mathcal{B}_{|\rho\in\Omega_{o}}}:= &\textcolor{blue}{\;\;\, B\int_{I} (\mathrm{d}_{\rho}\sigma) \overleftarrow{\mathbf{E}}^{2}\;} +\nonumber \\ &\textcolor{blue}{ +B\int_{I}\sum_{i \in \{1,2\}} 2(\mathrm{d}_{\rho}\kappa^{(i)}) \Re\left\{\frac{(\partial_{\tau}\overleftarrow{\mathbf{Q}}^{(i)})^2}{\varepsilon_{0}c^{(i)}}\right\}}.
\end{empheq}
\end{subequations}
From the derived equations, we can evaluate the procedure for two important special cases: 
\begin{itemize}
    \item If both background and design material in $\Omega$ do not contain any CCPR poles, i.e. they are both  non-dispersive and their losses are only described by the $\sigma$ term in Eq.~(\ref{Eq:CCPR_model}), all auxiliary fields vanish. And since \mbox{$\mathbf{S}_{\partial_{\tau}Q}^{(i)}\equiv0$}, no auxiliary adjoint source will be injected during the adjoint simulation. In addition, the gradients in Eqs.~(\ref{Eq:GradientsTimeReversal_Diss}) do not contain any terms related to the $\mathbf{Q}^{(i)}$ fields.
    \item If the objective function in Eq.~(\ref{Eq:ObjectiveEnerg&Dissipation}) defined on $\Omega_{o}$ does not depend on the density itself, i.e. the density does not vary in $\Omega_{o}$ during the optimization, the derivatives of the material parameters with respect to the density vanish in $\Omega_{o}$. And since $\forall\rho\in\Omega_{o}$: $\mathrm{d}_{\rho}\varepsilon_{\infty}=\mathrm{d}_{\rho}\sigma=\mathrm{d}_{\rho}\kappa^{(i)}\equiv0$, it follows that $\mathcal{B}_{|\rho\in\Omega_{o}}\equiv0$, i.e. the total gradient reduces to $\nabla_{\rho}F:= \mathcal{A}_{|\rho\in\Omega \setminus \Omega_o}$, in agreement with the result presented in Eq.~(\ref{Eq:Gradients1}).
\end{itemize}
\newpage
\subsubsection{\textsc{Extension to a diagonal, anisotropic permittivity with arbitrary number of poles}}
For the extension to the general case of an diagonal permittivity (including anisotropy) and arbitrary number of poles, the power dissipation density in (\ref{Eq:InstantanousDissipation}) must be extended to
\begin{equation}\label{Eq:InstantanousDissipation_with_poles}
q_e([\rho];t) = \sum_{k=1}^3\!\left(\sigma_k E_k^{2} + 2\!\!\sum_{i \in \{1,2\}} \sum_{p=1}^{P_{k}^{(i)}} \kappa^{(i)} \Re\left\{\frac{(\partial_{t}Q_{p,k}^{(i)})^2}{\varepsilon_{0}c_{p,k}^{(i)}}\right\}\!\right). 
\end{equation}
Accordingly, one must solve Eqs.~(\ref{Eqs:Dissipation}.b) for the auxiliary fields $\tilde{Q}_{p, k}^{(i)}$ for all poles $p\in\{1,...,P^{(i)}_{k}\}$ for the corresponding material $i=1,2$ and field component $k=1,2,3$ and sum over their derivatives in Eq.~(\ref{Eqs:Dissipation}.b), i.e. the adjoint system of Maxwell's equations in $\Omega_{o}$ reads
\begin{subequations}\label{Eqs:Dissipation_withpoles}
  \begin{empheq}[]{align}
-(\nabla \times \tilde{H})_k+\varepsilon_0 \varepsilon_{\infty,k} \partial_t \tilde{E}_k+\sigma_k \tilde{E}_k + \quad\nonumber \\  +2\sum_{i \in \{1,2\}} \sum_{p=1}^{P_{k}^{(i)}}\kappa^{(i)} \Re\left\{\partial_t \tilde{Q}_{p,k}^{(i)}\right\}&=\textcolor{Green}{S_{E_k}},\\[1pt]
\text{For } i =1,2 \text{ and } \forall p \in \{1, \ldots, P_k^{(i)}\}:  \nonumber \\ \partial_{\tau}\tilde{Q}_{p,k}^{(i)}-a_{p,k}^{(i)}\tilde{Q}_{p,k}^{(i)}- \varepsilon_{0}c_{p,k}^{(i)}\tilde{E}_k&=\textcolor{purple}{S_{\partial_{\tau}Q_{p,k}}^{(i)}},\\[8pt]
\mu_0 \partial_{\tau}\tilde{H}_{k} +(\nabla \times {\tilde{E}})_{k} &=0, 
\end{empheq}
\end{subequations}
with the adjoint source terms 
\begin{equation}\label{Eq:Adjoint_Sources_with_poles}
\begin{split}
\textcolor{Green}{S_{E_k}:=}& \textcolor{Green}{\phantom{-}2B\sigma_k\overleftarrow{E_k}},\\
\textcolor{purple}{S_{\partial_{\tau}Q_{p,k}}^{(i)}:=}&\textcolor{purple}{\phantom{-}2 B \partial_{\tau}\overleftarrow{Q}_{p,k}^{(i)}}, \;\, \text{for }i=1,2 \text{ and } \forall p \in \{1, \ldots, P_k^{(i)}\}.
\end{split}
\end{equation}
And the gradient terms from Eq.~(\ref{Eq:GradientsTimeReversal_Diss}) change to
\begin{subequations}\label{Eq:GradientsTimeReversal_Diss_with_poles}
\begin{empheq}[]{align}
& \mathcal{A}_{|\rho\in\Omega}:= \nonumber \\ 
& \quad \sum_{k=1}^{3}\Bigg(-\int_{I}\varepsilon_{0}(\mathrm{d}_{\rho}\varepsilon_{\infty,k})\partial_{\tau}\tilde{E}_k \cdot \overleftarrow{E}_k +\nonumber \\ 
&\quad  -\int_{I}(\mathrm{d}_{\rho}\sigma_k) \tilde{E}_k \cdot \overleftarrow{E}_k +\nonumber \\ 
&\quad   -\int_{I}\sum_{i \in \{1,2\}}\sum_{p=1}^{P_{k}^{(i)}}2(\mathrm{d}_{\rho}\kappa^{(i)})\partial_{\tau}\tilde{E}_k\cdot\Re\left\{\overleftarrow{Q}_{p,k}^{(i)}\right\}\Bigg), \\
& \textcolor{blue}{\mathcal{B}_{|\rho\in\Omega_{o}}}:= \nonumber \\ 
& \quad  \textcolor{blue}{\sum_{k=1}^{3}\Bigg( B\int_{I} (\mathrm{d}_{\rho}\sigma_k) \overleftarrow{E}_k^{2}+ }  \nonumber   \\
 & \quad   \textcolor{blue}{+B\int_{I}\sum_{i \in \{1,2\}}\sum_{p=1}^{P_{k}^{(i)}} 2(\mathrm{d}_{\rho}\kappa^{(i)}) \Re\left\{\frac{(\partial_{\tau}\overleftarrow{Q}_{p,k}^{(i)})^2}{\varepsilon_{0}c_{p,k}^{(i)}}\right\}\Bigg)}.
\end{empheq}
\end{subequations}

\subsubsection{\textsc{Discretization of the adjoint equations within the FDTD framework}}\label{sec:discretization}
\begin{figure*}
    \centering
    \includegraphics[width=0.8\linewidth]{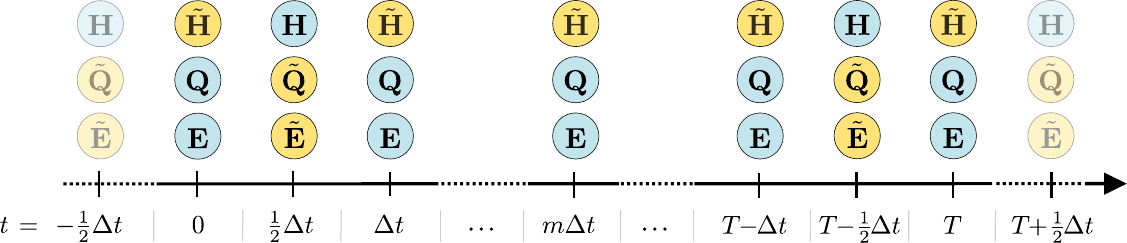}
    \caption{\raggedright Schematic illustration of the discretized (forward and adjoint) fields sampled on the discretized timeline within the FDTD framework.}
    \label{fig:DiscretizationSI}
\end{figure*}

In the FDTD framework, Maxwell's equations for both forward and adjoint simulation are discretized on a staggered \textit{Yee} grid in space and time \cite{Taflove}. For the discretization the time interval $[0,T]$ is considered to be divided into $M+1$ time steps $t_m:=\frac{m}{M}T=m\Delta t$, $m\in \{0,...,M\}$. We follow the same procedure presented in Ref.~\cite{material}. Here, the Eqs.~(\ref{Eq:Maxwells2}a) and (\ref{Eq:Maxwells2}b) for the forward fields are discretized at time step $(m + 1/2)\Delta t$, and Eq.~(\ref{Eq:Maxwells2}c) at time step $m\Delta t$. Consequently, the adjoint electric source term $S_{E_k}$ in Eq.~(\ref{Eq:Adjoint_Sources_with_poles}) is evaluated at time step $m\Delta t$, and the adjoint auxiliary source terms $S_{\partial_{\tau}Q_{p,k}}^{(i)}$ are sampled at time step $(m + 1/2)\Delta t$ while storing them during the forward simulation. To avoid a different update scheme for the adjoint system than for the forward system, we discretize both Eqs.~(\ref{Eqs:Dissipation_withpoles}.a) and (\ref{Eqs:Dissipation_withpoles}.b) at time step $m\Delta t$, and (\ref{Eqs:Dissipation_withpoles}.c) at time step $(m + 1/2)\Delta t$. Thus, both adjoint electric and auxiliary field must be sampled at time step $(m + 1/2)\Delta t$, and the adjoint magnetic fields must be sampled on time step $m\Delta t$. A schematic illustration of the disposition of the discretized fields on the timeline is demonstrated in Fig.~\ref{fig:DiscretizationSI}.\newline\newline
The updated equation for the adjoint auxiliary fields in Eqs.~(\ref{Eqs:Dissipation}b)  at time $m\Delta t$ reads 
\begin{equation}\label{Eq:Q_auxiliary_update}
\begin{split}
\tilde{Q}_{p,k}^{m+ \frac{1}{2}, (i)} & =\frac{2+a_{p,k}^{(i)} \Delta t}{2-a_{p,k}^{(i)} \Delta t} \tilde{Q}_{p,k}^{m - \frac{1}{2}, (i)} +\\ 
& + \frac{\varepsilon_0 c_{p,k}^{(i)} \Delta t}{2-a_{p,k}^{(i)} \Delta t}\left(\tilde{E}_{k}^{m+\frac{1}{2}}  + \tilde{E}_{k}^{m-\frac{1}{2}}\right) + \\ 
&+ \frac{2\Delta t}{2-a_{p,k}^{(i)} \Delta t}\textcolor{purple}{S_{\partial{\tau}Q_{p,k}}^{m, (i)}},
\end{split}
\end{equation}
where we approximate the auxiliary adjoint source term $S_{\partial{\tau}Q_{p,k}}^{m, (i)}$ at time step $m\Delta t$ using the centered difference approximation,
\begin{equation}
\begin{split}
S_{\partial{\tau}Q_{p,k}}^{m, (i)}&=2B\partial_{\tau}\overleftarrow{Q}_{p,k}^{(i)}\left[m\Delta t\right] \\
& =2B\;\frac{\left(Q_{p,k}^{M-m-1,(i)}-Q_{p,k}^{M-m+1,(i)}\right)}{2\Delta t}.
\end{split}
\end{equation}

The update of the adjoint electric field components are obtained via
\begin{equation}
\tilde{E}_{k}^{m+\frac{1}{2}}=\frac{\gamma_{k}^{m-\frac{1}{2}}}{\alpha_{k}},
\end{equation}
where $\alpha_k$ is a density-depended constant
\begin{equation}
\alpha_k:= \frac{\varepsilon_{0}\varepsilon_{\infty, k}}{\Delta t}  + \frac{\sigma_k}{2} +\frac{2}{\Delta t}\sum_{i \in \{1,2\}}\sum_{p=1}^{P_{k}^{(i)}}\kappa^{(i)}\Re\left\{\frac{\varepsilon_0 c_{p,k}^{(i)} \Delta t}{2-a_{p,k}^{(i)} \Delta t}\right\},
\end{equation}
and $\gamma_{k}^{m-\frac{1}{2}}$ changes at each time step according to
\begin{equation}\label{Eq:Gamma_Update_with_poles}
\begin{split}
& \gamma_{k}^{m-\frac{1}{2}}= \phantom{+}(\nabla \times \tilde{\mathbf{H}})_{k}^{m} +\\
&\!\!+\!\!\left(\! \frac{\varepsilon_{0}\varepsilon_{\infty,k}}{\Delta t}  \!-\! \frac{\sigma_k}{2} \!-\! \frac{2}{\Delta t}\!\!\!\!\sum_{i \in \{1,2\}} \!\sum_{p=1}^{P_{k}^{(i)}}\kappa^{(i)}\Re\!\left\{\frac{\varepsilon_0 c_{p,k}^{(i)} \Delta t}{2-a_{p,k}^{(i)} \Delta t}\right\}\!\!\right)\!\!\tilde{E}_{k}^{m-\frac{1}{2}} \\
&\!\!\!\! -\frac{2}{\Delta t}\sum_{i \in \{1,2\}} \sum_{p=1}^{P_{k}^{(i)}} \kappa^{(i)}\Re\left\{\frac{2a_{p,k}^{(i)} \Delta t}{2-a_{p,k}^{(i)} \Delta t}\tilde{Q}_{p,k}^{m-\frac{1}{2},(i)}\right\} + \\
&\textcolor{Green}{\!\!+ S^{m}_{E_k}}\;-\frac{2}{\Delta t}\sum_{i \in \{1,2\}} \sum_{p=1}^{P_{k}^{(i)}} \kappa^{(i)}\Re\left\{\frac{2\Delta t}{2-a_{p,k}^{(i)} \Delta t}\left({\color{purple} S_{\partial{\tau}Q_{p,k}}^{m, (i)}}\right) \right\}\\[10pt]
=&\phantom{+}(\nabla \times \tilde{\mathbf{H}})_{k}^{m} +  \\
&\!\!\!\!\!+\!\!\left(\! \frac{\varepsilon_{0}\varepsilon_{\infty,k}}{\Delta t}  \!-\! \frac{\sigma_k}{2}\! -\!\frac{2}{\Delta t}\!\!\sum_{i \in \{1,2\}}\!\sum_{p=1}^{P_{k}^{(i)}}\kappa^{(i)}\Re\left\{\frac{\varepsilon_0 c_{p,k}^{(i)} \Delta t}{2-a_{p,k}^{(i)} \Delta t}\right\}\!\!\right)\!\!\tilde{E}_{k}^{m-\frac{1}{2}} \\ 
& \!\!\!\!\!-\frac{2}{\Delta t}\sum_{i \in \{1,2\}} \sum_{p=1}^{P_{k}^{(i)}} \kappa^{(i)}\Re\left\{\frac{2a_{p,k}^{(i)} \Delta t}{2-a_{p,k}^{(i)} \Delta t}\tilde{Q}_{p,k}^{m-\frac{1}{2},(i)}\right\} +\\
&\textcolor{Green}{\!\!\!\!\!+\; 2B\sigma E_{k}^{M-m} } +\\ 
&\!\!\!\!\!-\frac{2}{\Delta t}\sum_{i \in \{1,2\}} \sum_{p=1}^{P_{k}^{(i)}} \kappa^{(i)}\Re\left\{\frac{2\Delta t}{2-a_{p,k}^{(i)} \Delta t}\left({\color{purple}2 B \partial_{\tau}Q_{p,k}^{M-m,(i)}}\right) \right\}.
\end{split}
\end{equation}
Based on the linear interpolation of design and background material as presented in Eq.~(\ref{Eq:InterpolatedModel}) while assigning a density value $\rho_{k,n}$ at the location of each electric field component $E_k$ in the $n$-th Yee cell, the gradients terms from Eqs.~(\ref{Eq:GradientsTimeReversal_Diss}) can then be calculated as 
\begin{equation}\label{Eq:Gradient_FDTD_Dissipation_with_poles}
 \begin{split}
&\mathcal{A}_{\rho_{k,n}\in\Omega}= \\ 
&\!\!\!-\!\,\varepsilon_{0}\left(\varepsilon_{\infty, k,n}^{(2)} - \varepsilon_{\infty, k,n}^{(1)}\right)\!\!\sum^{M}_{m=0}E^{M-m}_{k,n}\left(\tilde{E}^{m + \frac{1}{2}}_{k,n} - \tilde{E}^{m - \frac{1}{2}}_{k,n}\right)+ \\
& \!\!\!-\!\!\left(\sigma_{k,n}^{(2)} \!-\! \sigma_{k,n}^{(1)} \!+\! (1\!-\!2\rho_{k,n})\gamma\right)\!\!\sum^{M}_{m=0}\!E^{M-m}_{k,n}\frac{\left(\tilde{E}^{m + \frac{1}{2}}_{k,n} \!+\! \tilde{E}^{m - \frac{1}{2}}_{k,n}\right)}{2}\,\Delta t +\\
& \!\!-2\!\!\sum^{M}_{m=0}\sum_{i \in \{1,2\}}\!\sum_{p=1}^{P_{k}^{(i)}}\partial_{\rho}\kappa^{(i)}\Re\left\{Q^{M-m, (i)}_{p, k, n}\right\}\left(\tilde{E}^{m + \frac{1}{2}}_{k,n} - \tilde{E}^{m - \frac{1}{2}}_{k,n}\right),  
\end{split} 
\end{equation}

\begin{equation}\label{Eq:Gradient_FDTD_Dissipation_with_polesb}
 \begin{split}
&\textcolor{blue}{\mathcal{B}_{\rho_{k,n}\in\Omega_{o}}}= \\
& \textcolor{blue}{\!\!+B\left(\sigma_{k,n}^{(2)} - \sigma_{k,n}^{(1)} + (1-2\rho_{k,n})\gamma\right)\sum^{M}_{m=0}\left(E^{M-m}_{k,n}\right)^2\Delta t} +\\
& \textcolor{blue}{\!\!+2B\!\!\sum^{M}_{m=0}\sum_{i \in \{1,2\}}\sum_{p=1}^{P_{k}^{(i)}}\partial_{\rho}\kappa^{(i)}\Re\!\left\{\!\!\frac{\left(Q^{M-1-m, (i)}_{p, k, n} \!-\! Q^{M-m+1, (i)}_{p, k, n}\right)^2 \!\!}{4\,\Delta t\,\varepsilon_{0}c^{(i)}_{p,k}}\right\}}
\end{split} 
\end{equation}
where $\tilde{E}^{-\frac{1}{2}}_{k,n}=Q^{M+1, (i)}_{k, n}=0$ according to the initial and boundary conditions respectively. Using this equation, we find that filtering the density values $\rho_{k,n}$ is essential, as otherwise the density might converge to three different component-wise structures that do not result in a physically meaningful topology. An extraction method of the final design based on a density distribution on the staggered Yee grid, is presented in Ref~\cite{Gedeon2023}.

\bibliographystyle{IEEEtran.bst}
\bibliography{bibliography.bib}